\documentclass[reprint, amsmath,amssymb,showpacs,floatfix, aps, prb, twocolumn, superscriptaddress]{revtex4-2}
\usepackage{graphicx}
\usepackage{dcolumn}
\usepackage{bm}
\usepackage{color}
\usepackage{hyperref}
\hypersetup{colorlinks=true,citecolor=blue,linkcolor=blue, urlcolor=blue}
\newcolumntype{M}[1]{>{\centering\arraybackslash}m{#1}}
\newcolumntype{N}{@{}m{0pt}@{}}
\def\nn{\nonumber}
\def\({\left(}
\def\){\right)}
\def\[{\left[}
\def\]{\right]}

\def\b{\beta}
\def\d{\delta}

\def\e{\epsilon}

\def\s{\sigma}
\def\t{\tau}
\def\w{\omega}
\def\v{\nu}

\def\D{\Delta}

\def\bfk{\mathbf{k}}

\def\bfq{\mathbf{q}}
\def\bfr{\mathbf{r}}

\def\>{\rangle}
\def\<{\langle}
\def\bfG{\mathbf{G}}

\def\dr{d^3 {\mathbf{r}}}
\def\drp{d^3 {\mathbf{r}'}}

\begin{document}

\title{Development of $ab ~initio$ method for exciton condensation and its application to $\bf TiSe_2$}

\author{Hsiao-Yi Chen}
\affiliation {RIKEN Center for Emergent Matter Science (CEMS), Wako 351-0198, Japan}

\author{Takuya Nomoto}
\affiliation {Research Center for Advanced Science and Technology,
University of Tokyo, Komaba Meguro-ku, Tokyo 153-8904, Japan}

\author{Ryotaro Arita}
\affiliation {RIKEN Center for Emergent Matter Science (CEMS), Wako 351-0198, Japan}
\affiliation {Research Center for Advanced Science and Technology,
University of Tokyo, Komaba Meguro-ku, Tokyo 153-8904, Japan}

\date{\today}

\begin{abstract}
Exciton condensation indicating the spontaneous formation of electron-hole pair can cause the phase transition from a semimetal to an excitonic insulator by gap opening at the Fermi surface. 
While the idea of this excitonic insulator has been proposed for decades, current theoretical approaches can only provide qualitative descriptions, and a quantitative predicting tool is still missing. 
To shed insight on this problem, we developed an {\it ab initio} method based on the finite-temperature density functional theory and many-body perturbation theory to compute the exciton condensation critical behavior. Applying our approach to the monolayer $\rm TiSe_2$, we find a lattice distortion accompanied by the formation of the excitonic gap via electron-phonon coupling without phonon softening, proving that the exciton condensation is the origin of the charge-density-wave state observed in this compound. Overall, the methodology introduced in this work is general and paves the way to searching for candidate excitonic insulators in natural material systems.
\end{abstract}
\maketitle
\section{Introduction}
\vspace{-10pt}
The formation of fermion pairs can break the limit of the Pauli exclusion principle, accepting particles to occupy the same quantum state. At low temperatures, the coherent occupation can reshape the electronic structure and lead to a phase transition. The most well-known example in metallic materials is the Bardeen-Cooper-Schrieffer (BCS) theory of superconductivity \cite{bardeen1957microscopic} where the paired electrons, the Cooper pairs, open a gap that supports frictionless transport and provoke superconductivity. In a semimetal or a semiconductor with a small gap, on the other hand, a similar mechanism pairs up an electron in the conduction band with a hole in the valence band to form the so-called exciton, and the condensation of excitons in the ground state gives rise to the excitonic insulator (EI) \cite{mott1961transition,keldysh1965possible,Jerome1967exciton,halperin1968possible}.
\\
\indent
The theoretical description for EI is so far well-established under different pairing conditions, ranging from weak coupling BCS-like scenario \cite{Jerome1967exciton} to strongly coupling Bose-Einstein condensation \cite{bronold2006possibility,tomio2006excitonic,zenker2012electron}, and with pairing force beyond the Coulomb attraction \cite{phan2013exciton,de2018interaction,murakami2017photoinduced}.
Methods are proposed as an analogy of superconductivity theory by replacing the superconducting order parameter $\chi=\<\psi_e\psi_e\>$ with the thermal expectation value of exciton operator $\chi^{\rm ex}=\<\psi_e\psi_h\>$. 
However, decades after the idea came out, the progress in experimental searching for an excitonic insulator is still way behind. Currently, the exciton condensation is mainly realized in artificial quantum well and bilayer structure, or induced by pressure and photo-excitation \cite{
neuenschawander1990pressure,bucher1991excitonic,eisenstein2004bose,li2016negative,du2017evidence,gupta2020heterobilayers,ma2021strongly,okazaki2018photo}. On the other hand, natural cases are only found in pristine crystals of some transition metal compounds \cite{seki2014excitonic,hedayat2019excitonic,lu2017zero,sugimoto2018strong,kim2021direct,kogar2017signatures}.
\\
\indent
Among them, 1$T$-$\rm TiSe_2$ in the bulk form is observed to undergo structural phase transition from a disorder state to a commensurate $2\times 2\times 2$ charge-density-wave (CDW) state when the temperature is decreased below the critical value $T_c\sim 190~$K \cite{hellmann2012time,kidd2002electron,weber2011electron,kogar2017signatures,rossnagel2002charge,
cercellier2007evidence}. The phase transition is characterized by the mixing of the Se $4p$ and Ti $3d$ bands observed by ARPES \cite{rossnagel2002charge,cercellier2007evidence} and the occurrence of phonon softening
revealed by X-ray diffuse scattering \cite{holt2001x}. However, these observations are insufficient and sometimes controversial to understand the fundamental mechanism to form the CDW order, for which theorists argued between the band Jahn-Teller effect \cite{hughes1977structural,suzuki1985microscopic,whangbo1992analogies} and exciton condensation \cite{wilson1978infrared,pilo2000Photoemission,cercellier2007evidence,monney2009spontaneous,van2010exciton,rossnagel2011origin,zenker2013chiral,kogar2017signatures,
hellgren2017critical,hedayat2019excitonic}.
The puzzle lasted until the recent experiment using the momentum-resolved electron energy-loss (EEL) spectroscopy \cite{kogar2017signatures}, where a soft-plasmon mode that appears exclusively in the EI phase was observed. In recent years, with the development of the two-dimensional synthesis technique, 1$T$-$\rm TiSe_2$ in its monolayer form is raising a new trend to study the EI/CDW properties \cite{chen2015charge,sugawara2016unconventional,singh2017stable,duong2017raman,
fang2017x,kolekar2017layer,kaneko2018exciton} and its relation to superconductivity \cite{li2016controlling}.
\\
\indent
In contrast to the versatile model development and bountiful experimental evidence, numerical methods to compute EI from first principles are remarkably lacking. 
Current approaches using density functional theory (DFT)  based on the local exchange-correlation (XC) function such as the local density approximation (LDA), generalized gradient approximation (GGA), or hybrid functional \cite{ceperley1980ground,perdew1996generalized,heyd2003hybrid} can describe the ground state structure of ordered and disorder phases \cite{pasquier2018excitonic,guster20182,varsano2020monolayer,kaneko2012electronic} but not the evolution as a function of temperature. 
A rare example can be found with the assistance of the GW and Bethe-Salpeter equation (BSE) method to construct the Bogoliubov-de-Gennes equation and calculate the EI gap function at finite temperature \cite{varsano2017carbon}. However, the approach employs only the Coulomb attraction, and the interplay between electron density and the crystal structure is missed.
\\
\indent
In this paper, we promote the EI order parameter to a generalized nonlocal (NL) density and construct a DFT scheme for exciton condensation, as inspired by the superconductivity density functional theory (SCDFT) \cite{kurth1999local,luders2005ab}. For the lattice degree of freedom, we adopt the Born-Oppenheimer approximation with additional linear potential and take account of the anharmonicity to study the possible atomic displacement in the ordered phase. 
Treating the electron and phonon on equal footing, we use the many-body perturbation method to compute the self-consistent NL XC functional and derive the gap equation to investigate the critical behavior. Based on the formalism, we carry out the numerical implementation for monolayer $\rm TiSe_2$ and discuss the determining component to calculate the transition temperature. We show that the underlying mechanism causing the formation of the CDW order in monolayer $\rm TiSe_2$ is the strong electron-phonon coupling mixing the occupied and unoccupied states. Our work presents a broadly applicable approach to access the microscopic mechanism of exciton condensation and shed light on the numerical method to reveal the EI phase in materials.
\\
\indent
The paper is organized as follows. In Sec.~\ref{sect:theory} we develop our $ab~initio$ method to describe exciton condensation and provide a brief review on phonon self-energy. In Sec.~\ref{Sect:ex-cond22} we derive the gap equation for the EI/CDW state caused by electron-phonon (e-ph) coupling. In Sec.~\ref{Sect:ex_TiSe2} we applied the formalism and carry out numerical calculation to study the EI/CDW phase in monolayer $\rm TiSe_2$. We summarize the results and discuss future research in Sec.~\ref{Sect:conclusion}. In the Appendices, We left derivations and numerical details in the appendices.
\section{Theory \label{sect:theory}}
\vspace{-10pt}
This section explains the formulation involved in this work. We present a DFT formalism with a general NL density characterizing the spontaneous fermion pairing, where the XC functional and ground state structure can be obtained self-consistently at finite temperature by the thermodynamical method. Then we extend the discussion by including the phonon degree of freedom. A linear distortion potential is introduced to study CDW. Before processing, we digress to summarize phonon self-energy and the self-consistent phonon (SCP) theory \cite{tadano2018first} which is dominated by the anharmonic phonon-phonon interaction. Overall, the theories adopted and developed here will be applied to investigate exciton condensation from first principles in the following section.
\subsection{DFT with generalized nonlocal density \label{subsect:DFT-NL}}
\vspace{-10pt}
The {\it ab initio} approach adopted in this work is based on the multicomponent DFT \cite{kreibich2001multicomponent}. By generalizing the electronic density, this approach has opened the gate to computing molecule bonding length, magnetization, and superconductivity from first principles \cite{kreibich2001multicomponent,von1972local,oliveira1988density}. Starting from the many-electron interacting Hamiltonian 
\begin{equation}
\label{Eq:interacting-H}
\hat{H}=\hat{T}+\hat{U}^{\rm ee}+\hat{V}_{{\rm ext}}
\end{equation}
where the $\hat{T}$ is the electronic kinetic energy, $\hat{U}^{\rm ee}$ is the Coulomb interaction among electron gas, and  $\hat{V}_{\rm ext}$ is the external field including the potential from the nuclei. The Kohn-Sham (KS) method \cite{kohn1965self} replaces the many-body interaction by a local XC potential and writes down an effective single-particle Hamiltonian:
%
\begin{equation}
\label{Eq:KS-H}
\hat{H}^{{ \rm KS}}=\hat{T}[n]+\hat{V}_{{\rm XC}}[n]
\end{equation}
which can reproduce the same ground state energy as  Eq.~(\ref{Eq:interacting-H}) with the same local electronic density according to the Hohenberg-Kohn theorem \cite{hohenberg1964inhomogeneous}. Thus, if we have the true XC potential, we can obtain the exact solution for the target system, but $\hat{V}_{\rm XC}$ can only be acquired approximately. 
\\
\indent
In correlated systems, like superconductor \cite{oliveira1988density,luders2005ab}, the local potential is insufficient to represent the many-body interaction, and NL potential and density must be adopted to characterize the electronic order. Here, we start from a DFT with the local charge density $n(\bfr)$ and the NL charge densities $\chi(\bfr,\bfr')$ in the form:
\begin{align}
&n(\bfr)=\sum_{\sigma}\<\Psi_\s^\dagger(\bfr)\Psi_\s(\bfr)\>\nn\\
&\chi(\bfr,\bfr')=\sum_{\sigma}\<\Psi_\s^\dagger(\bfr)\Psi_{\s}(\bfr')\>
\label{Eq:def-chi}
\end{align}
where $\Psi_\s$ is the electron operator, $\s$ denotes the spin index, and $\<A\>={\rm Tr}\left[e^{-\b H}A\right]$ defines the thermal expectation value. Therefore, following the KS construction of DFT, we can include possible external potential acting on the NL charge density by considering the Hamiltonian:
\begin{equation}
\label{Eq:interacting-NLH}
\hat{H}=\hat{T}+\hat{U}^{\rm ee}+\hat{V}_{\rm ext}+\hat{\Delta}_{\rm ext}
\end{equation}
where
\begin{align}
&\hat{V}_{\rm ext}=\sum_{\s}\int \dr\Psi^\dagger_\s(\bfr)v_{\rm ext}(\bfr)\Psi_\s(\bfr)\nn\\
&\hat{\D}_{\rm ext}=\sum_{\s}\int \dr\Psi^\dagger_\s(\bfr)\Delta_{\rm ext}(\bfr,\bfr')\Psi_{\s}(\bfr').
\end{align}
At finite temperature, this Hamiltonian reproduces the thermodynamic potential as a functional of local and NL charge density:
\begin{align}
\Omega[n,\chi]=& F[n,\chi]+\int \dr~ n(\bfr)\[v_{\rm ext}(r)-\mu\]\nn\\
&+\int \dr \drp\[\chi(\bfr,\bfr')\D_{\rm ext}(\bfr,\bfr')+h.c.\]
\end{align}
where the free energy $F[n,\chi]$ is universal and independent from external potentials.
\\
\indent
Compared to the interacting Hamiltonian, Eq.~(\ref{Eq:interacting-NLH}), we can write down the corresponding KS Hamiltonian of non-interacting orbitals \cite{kohn1965self}:
\begin{equation}
\label{Eq:KS-NLH}
\hat{H}^{\rm KS}=\hat{T}+\hat{V}_{\rm XC}+\hat{\Delta}_{\rm XC}
\end{equation}
and the thermodynamic potential becomes:
\begin{align}
\Omega^{\rm KS}[n,\chi]&=F^{\rm KS}[n,\chi]+F^{\rm L}_{\rm XC}[n]+F^{\rm NL}_{\rm XC}[\chi]
\nn\\
&+\frac{1}{2}\int \dr\drp \frac{n(\bfr)n(\bfr')}{|\bfr-\bfr'|}
-\mu\int \dr n(\bfr)
\end{align}
where $F^{\rm KS}[n,\chi]$ is the free energy of non-interacting KS system, and $F^{\rm L}_{\rm XC}[n]$ and $F^{\rm NL}_{\rm XC}[\chi]$ are the XC free energy from the local and  NL densities. To make the KS Hamiltonian Eq.~(\ref{Eq:KS-NLH}) reproduce the same thermal ground state as the interacting system, both thermodynamic potentials must be minimized by the same ground state densities:
\begin{align}
\label{Eq:dOdn}
&\frac{\partial\Omega}{\partial n(\bfr)}=\frac{\partial F}{\partial n(\bfr)}+v_{\rm ext}(\bfr)-\mu=0\nn\\
&\frac{\partial\Omega^{\rm KS}}{\partial n(\bfr)}=
\frac{\partial F^{\rm KS}}{\partial n(\bfr)}+
\frac{\partial F^{L}_{\rm XC}}{\partial n(\bfr)}
+\int \drp \frac{n(\bfr')}{|\bfr-\bfr'|}-\mu=0
\end{align}
and
\begin{align}
\label{Eq:dOdchi}
&\frac{\partial\Omega}{\partial \chi(\bfr,\bfr')}=\frac{\partial F}{\partial \chi(\bfr,\bfr')}+\D_{\rm ext}(\bfr,\bfr')=0\nn\\
&\frac{\partial\Omega^{\rm KS}}{\partial \chi(\bfr,\bfr')}=
\frac{\partial F^{\rm KS}}{\partial \chi(\bfr,\bfr')}+
\frac{\partial F^{\rm NL}_{\rm XC}}{\partial \chi(\bfr,\bfr')}=0.
\end{align}
By defining the exchange-correlation potential:
\begin{align}
& v_{\rm xc}[n](\bfr)\equiv \frac{\partial F^{L}_{\rm XC}}{\partial n(\bfr)}\nn\\
& \D_{\rm xc}[\chi](\bfr,\bfr')\equiv \frac{\partial F^{\rm NL}_{\rm XC}}{\partial \chi(\bfr,\bfr')},
\end{align}
we can write down the generalized KS equation:
\begin{align}
\label{Eq:KS-D}
&\[-\frac{\nabla^2}{2}+\int \drp\frac{n(\bfr')}{|\bfr-\bfr'|}+v_{\rm xc}-\mu\]\psi_i(\bfr)\nn\\
&~~~~~+\int \drp \D_{\rm xc}(\bfr,\bfr')\psi_i(\bfr')=E_i\psi_i(\bfr)
\end{align}
The result shows that if we can obtain the exact form of the XC free energy, we can construct the exact ground state of the interacting system based on the KS theorem. This derivation is general for all NL-densities beyond the definition in Eq.~(\ref{Eq:def-chi}). For example, by choosing $\chi(\bfr,\bfr')=\<\Psi_{\uparrow}(\bfr)\Psi_{\downarrow}(\bfr')\>$, i.e. the Cooper pairing, we can obtain the DFT for superconductivity \cite{oliveira1988density,luders2005ab}.
\\
\indent
In this work, we apply the KS perturbation theory \cite{gorling1994exact} following the similar approach of SCDFT \cite{kurth1999local,luders2005ab} to determine the $F_{\rm XC}$'s. By identifying the KS Hamiltonian, Eq.~(\ref{Eq:KS-NLH}), as the unperturbed part from the interacting Hamiltonian, Eq~(\ref{Eq:interacting-NLH}), we obtain a perturbation theory:
\begin{equation}
\label{Eq:H_pert1}
\hat{H}_0=\hat{H}^{\rm KS}~~;~~\hat{H}_I=\hat{U}^{\rm ee}-\hat{V}_{\rm XC}-\hat{\D}_{\rm XC}.
\end{equation}
Provided with all order contributions in the $\hat{\D}_{\rm XC}$ operator, the difference between the interacting and KS thermodynamic potential can be expanded by the linked-cluster theorem \cite{mahan2013many} as:
\begin{align}
\label{Eq:Omega-Omegaks}
\Omega=\Omega^{\rm KS}-\frac{1}{\beta}\sum_{l=1}^\infty U_l
\end{align}
where $U_l$ are different connected Feynman diagrams. According to Eq.~(\ref{Eq:dOdn}) and Eq.~(\ref{Eq:dOdchi}), both thermal potentials are minimized by the ground state density so that the derivative of Eq.~(\ref{Eq:Omega-Omegaks}) must vanish and provide the condition to solve for the XC potential. In Appendix~\ref{Append:GW}, we show that this method can also be applied to derive the GW correction, which is commonly obtained from the Green's function method \cite{hybertsen1986electron}.
\subsection{DFT with generalized density from phonon degree of freedom}
\vspace{-10pt}
To discuss the structural transition, we must adopt a theory involving the atomic degree of freedom. In this work, we restrict the discussion to the Born-Oppenheimer approximation which uses the phonon vibration to represent the atomic motion. The standard approach expands the interatomic potential from the equilibrium structure in the Taylor series, where the lowest quadratic term defines the non-interacting phonon normal mode while higher-order terms take account of the anharmonic phonon-phonon (ph-ph) interactions \cite{baroni2001phonons}. As a result, the spontaneous lattice distortion can be characterized by the thermal average of the phonon operator, which is proportional to the static atomic displacement. 
\\
\indent
To accommodate the lattice deformation, we extend the KS Hamiltonian by introducing a phonon Hamiltonian:
\begin{equation}
\label{Eq:ph}
\hat{H}_{\rm ph}=\hat{H}_{\rm 0,ph}+\hat{U}^{\rm ph}_{\rm anh.}+\hat{U}^{\rm e-ph}+\hat{\D}_b
\end{equation}
where the free phonon Hamiltonian $\hat{H}_{\rm 0,ph}$, ph-ph anharmonic interaction $\hat{U}^{\rm ph}_{\rm anh.}$, and e-ph interaction $\hat{U}^{\rm e-ph}$ are defined as standard forms found in literature \cite{tadano2018first,bernardi2016first}. The last term denotes a linear deformation potential of the form:
\begin{equation}
\label{Eq:Db}
\hat{\D}_b=\sum_{\bfq\v}
\(\D_{b,\bfq\v}^*\hat{b}_{\bfq\v}+\D_{b,\bfq\v}\hat{b}^\dagger_{\bfq\v}\),
\end{equation}
which can push atoms away from their equilibrium position without phonon softening. Therefore, the phonon operator acquires a finite thermal average:
\begin{equation}
\label{Eq:chi_b}
    \<\hat{b}_{\bfq\nu}\>=\chi_{b,\bfq\nu}
\end{equation}
such that if we treat $\chi_{b,\bfq\nu}$ on the same footing as the densities in Eq~.(\ref{Eq:def-chi}), the $\D_b$'s can be determined by the perturbation expansion method introduced in the previous section where the Hamiltonian now includes the phonon degree of freedom:
\begin{subequations}
\begin{align}
&\hat{H}=\hat{H}_0+\hat{H}_{\rm I} \label{Eq:main-Ha}
\\
&\hat{H}_0=\hat{T}+\hat{V}_{\rm XC}+\hat{\D}_{\rm XC}+
\hat{H}_{\rm 0,ph}+\hat{\D}_b\label{Eq:main-Hb}
\\
&\hat{H}_{\rm I}=\hat{U}^{\rm ee}-\hat{V}_{\rm XC}-\hat{\D}_{\rm XC}+\hat{U}^{\rm ph}_{\rm anh.}+\hat{U}^{\rm e-ph}-\hat{\D}_b.\label{Eq:main-Hc}
\end{align}
\end{subequations}
This is the general formalism to describe lattice deformation at finite temperature. The distorted pattern follows the phonon normal vector denoted by the mode index $\nu$, and
the displacement amplitude is proportional to the absolute value of $\chi_{b,\bfq\nu}$. The momentum $\bfq$ will determine the periodicity of the new crystal structure. In Sec. \ref{Sect:ex-cond22} we will apply Eq.~(\ref{Eq:main-Ha}-\ref{Eq:main-Hc}) to derive the gap equation for the exciton condensation. 
\subsection{Phonon Self-energy from first-principle \label{subsect:ph_self}}
\vspace{-10pt}
In this section, we provide a discussion concerning the phonon properties obtained by the first-principles approaches. The {\it ab initio} phonon vibration can be calculated by diagonalizing the dynamical matrix, which encodes the total energy change when atoms move away from their equilibrium position. In practical approach, the dynamical matrix is computed by density functional perturbation theory (DFPT) \cite{baroni2001phonons} or using the frozen phonon method \cite{yin1980microscopic}, which gives the phonon vibration at zero temperature. 
Phonon frequencies obtained by these methods contain the "bare" part as well as the renormalization effect:
\begin{subequations}
\begin{align}
&\omega^2_{\bfq\nu}={\omega^{\rm bare}_{\bfq\nu}}^2+2\omega^{\rm bare}_{\bfq\nu} \Pi^{\rm bare}_{\bfq\nu}(T)
\label{Eq:ph_self_w}
\\
&\Pi^{\rm bare}_{\bfq\nu}(T)=\frac{2}{\mathcal{N}_k}\sum_{\bfk mn}
|g^{\rm bare}_{mn \nu }(\bfk,\bfq)|^2\frac{f_{\bfk+\bfq,m}-f_{\bfk,n}}{\xi^{m}_{\bfk+\bfq}-\xi^n_{\bfk}}\label{Eq:ph_self_Pi}
\end{align}
\end{subequations}
where $g^{\rm bare}_{mn\nu}(\bfk,\bfq)$ is the bare e-ph matrix element:
    \begin{equation}
    \label{Eq:eph_bare}
    g^{\rm bare}_{mn\nu}(\bfk,\bfq)=\(\frac{\hbar}{2\w_{\bfq\v}^{\rm bare}}\)^{1/2} \<\psi_{m\bfk+\bfq}|\partial_{\bfq \nu} V|\psi_{n\bfk}\>,
\end{equation}
and $\Pi^{\rm bare}_{\bfq\nu}$ is the self-energy due to the Coulomb screening, where the superscript "bare" denotes that we used the bare coupling constant from Eq.~(\ref{Eq:eph_bare}). The factor of 2 in Eq.~(\ref{Eq:ph_self_Pi}) appears for the spin degeneracy. $\mathcal{N}_k$ is the number of k-points, $\xi^n_{\bfk}$ is the electron energy of the $n$-th band with momentum $\bfk$ measured from the chemical potential, $f_{\bfk,n}=1/(e^{\b \xi^n_{\bfk}}+1)$ is the Fermi-Dirac distribution function.
Here we note that although the self-energy depends on the phonon frequency via the coupling constant, the renormalization only depends on the deformation potential $\partial_{\bfq \nu} V$ and the electronic occupations $f_{\bfk,n}$. 
Therefore, for later use, we define a frequency-independent factor 
\begin{equation}
\label{Eq:wPi}
X_{\bfq\nu}(T)=
\frac{\hbar}{\mathcal{N}_k}
\sum_{\bfk mn}
|\<\psi_{m\bfk+\bfq}|\partial_{\bfq \nu} V|\psi_{n\bfk}\>|^2
\frac{f_{\bfk+\bfq,m}-f_{\bfk,n}}{\xi^m_{\bfk+\bfq}-\xi^n_{\bfk}},
\end{equation}
such that the phonon frequency difference between two temperatures due to the screening effect can be written as
\begin{equation}
    \omega^2_{\bfq\nu}(T_1)-\omega^2_{\bfq\nu}(T_2)= 2X_{\bfq\nu}(T)\Big|^{T_1}_{T_2},
\end{equation}
where we neglect the temperature dependence in the deformation potential.
\\
\indent
In general, since experimental observables are the fully renormalized quantities, researchers focus on the explanation of the direct result $\omega$ as the experiment measurement, while the isolation between the bare frequency $\omega^{\rm bare}$ and self-energy is rarely studied \cite{berges2020ab}. However, when a single phonon momentum $\bfq$ connects the electron pocket and hole pocket in some metallic systems, $\Pi_{\bfq\nu}$ tends to diverge, which will force the corresponding renormalized phonon becomes soft with imaginary frequency since the self-energy is always negative. This so-called Fermi-surface nesting effect will destabilize the crystal structure, result in lattice distortion, and cause the transition into a CDW phase.
\\
\indent
Phonons derived from the above method are limited to the harmonic approximation, which expands the energy variation to the second order of atomic displacement. Although the harmonic phonon approach has great success in studying transport properties, carrier relaxation, and polaronic system \cite{bernardi2014ab,lee2018charge,Zhou2019predicting}, it fails to describe many important properties related to the lattice anharmonicity, such as thermal expansion, lattice transition, the temperature dependence of phonon frequency, and prediction of superconductivity critical temperature \cite{tadano2014anharmonic,tadano2015self,sano2016effect,
zhou2018electron,wang2022optimal,masuki2022ab}. 
\\
\indent
In this work, we adopt the self-consistent phonon (SCP) theory \cite{mauri2014anharmonic,tadano2018first} to compute the anharmonic phonon frequency. The SCP theory expands the energy to quartic order of the atomic displacement:
\begin{equation}
U_n=\frac{1}{n!}\(\frac{\hbar}{2}\)^{\frac{n}{2}}
\sum_{\{q_n\}} \d({\Sigma\bfq_n,\mathbf{G} })
\frac{\Phi(q_1,\cdots q_n)}{\sqrt{\w_{q_1}\cdots\w_{q_n}}}
\hat{A}_{q_1}\cdots \hat{A}_{q_n}
\end{equation}
where $n$ is the number of moved atom, $q=(\bfq,\v)$ is the collective index, $\d({\Sigma\bfq_n,\mathbf{G} })$ impose the momentum conservation, $\Phi(q_1,\cdots q_n)$ is the interatomic force constant, and $\hat{A}_{q}$ is the displacement operator. To the lowest order, the self-energy comes from the loop diagram of the 4-points vertex:
\begin{equation}
\label{Eq:Sigma_ph-ph}
\Sigma_{\bfq,\v\v'}=-\frac{1}{2}\sum_{q_1}
\frac{\hbar\Phi(\bfq \v,-\bfq \v',q_1,-q_1)}{4\sqrt{\w^{\rm scp}_{\bfq\v}\w^{\rm scp}_{-\bfq\v'}}\w^{\rm scp}_{q_1}}\times
(1+2n_{{\rm B}}(\w^{\rm scp}_{q_1}))
\end{equation}
where $n_{\rm B}$ is the Bose-Einstein distribution function. This anharmonic self-energy contribution contains the mixing between different phonon normal modes $\v, \v'$ such that it modifies both the phonon frequency and the vibration pattern. When the off-diagonal term can be neglected, the SCP equation can be simplified as:
\begin{align}
\label{Eq:w_scp}
{\w^{\rm scp}_{\bfq\nu}}^2=\omega^2_{\bfq\nu}+2\w^{\rm scp}_{\bfq\nu}\Sigma_{\bfq,\v\v}
\end{align}
It has been shown that the SCP equation requires a real solution for all $\Omega$'s such that the self-consistent solution breaks down when any phonon becomes soft.
\section{Exciton condensation \label{Sect:ex-cond22}}
\vspace{-10pt}
We apply the methods introduced in the previous section to study the EI/CDW phase. We will focus on the exciton formation between electron and hole with momentum $(\bfk_e,\bfk_h)=(\bfk,\bfk+{\bf M})$ where the transition momentum {\bf M} is chosen as $\mathbf{G}/2$, half of the reciprocal vector. For each wavevector $\bfk$, the eigenfunction of the generalized KS equation, Eq.~(\ref{Eq:KS-D}) can be written as the mixing of the Bloch-wave function in the disordered state:
\begin{equation}
\label{Eq:tphi}
\phi^{n}_{(\bfk)}=\sum_v t^{n}_{v\bfk} \psi_{v\bfk}+\sum_{c}t^{n}_{c\bfk+{\bf M}} \psi_{c\bfk+{\bf M}}=\sum_i t^n_{i(\bfk)}\psi_{i(\bfk)}
\end{equation}
where we use $v$ to denote the band index for states with momentum $\bfk$, $c$ to denote the band index for states with momentum $\bfk+{\bf M}$, and $(\bfk)$ for the mixed momentum $\bfk$ and $\bfk+{\bf M}$, and make the notation as a convention throughout the paper. On the other hand, the NL density becomes:
\begin{subequations}
\begin{align}
\label{Eq:NL-chi_a}
&{\chi}(\bfr,\bfr')=\sum_{ij(\bfk)}{\chi}^{ij}_{(\bfk)}\psi^{*}_{i(\bfk)}(\bfr)\psi_{j(\bfk)}(\bfr')
\\
&{\chi}^{ij}_{(\bfk)}
\equiv\sum_{n}f_{n(\bfk)}
t_{i(\bfk)}^{n*}
t_{j(\bfk)}^{n}
\label{Eq:NL-chi_b}
\end{align}
\end{subequations}
where $\chi^{ij}_{(\bfk)}$ evaluates the mixing between the $i$-th and $j$-th orbitals. The corresponding potential can be written as:
\begin{figure}[t]
\includegraphics[scale=0.25]{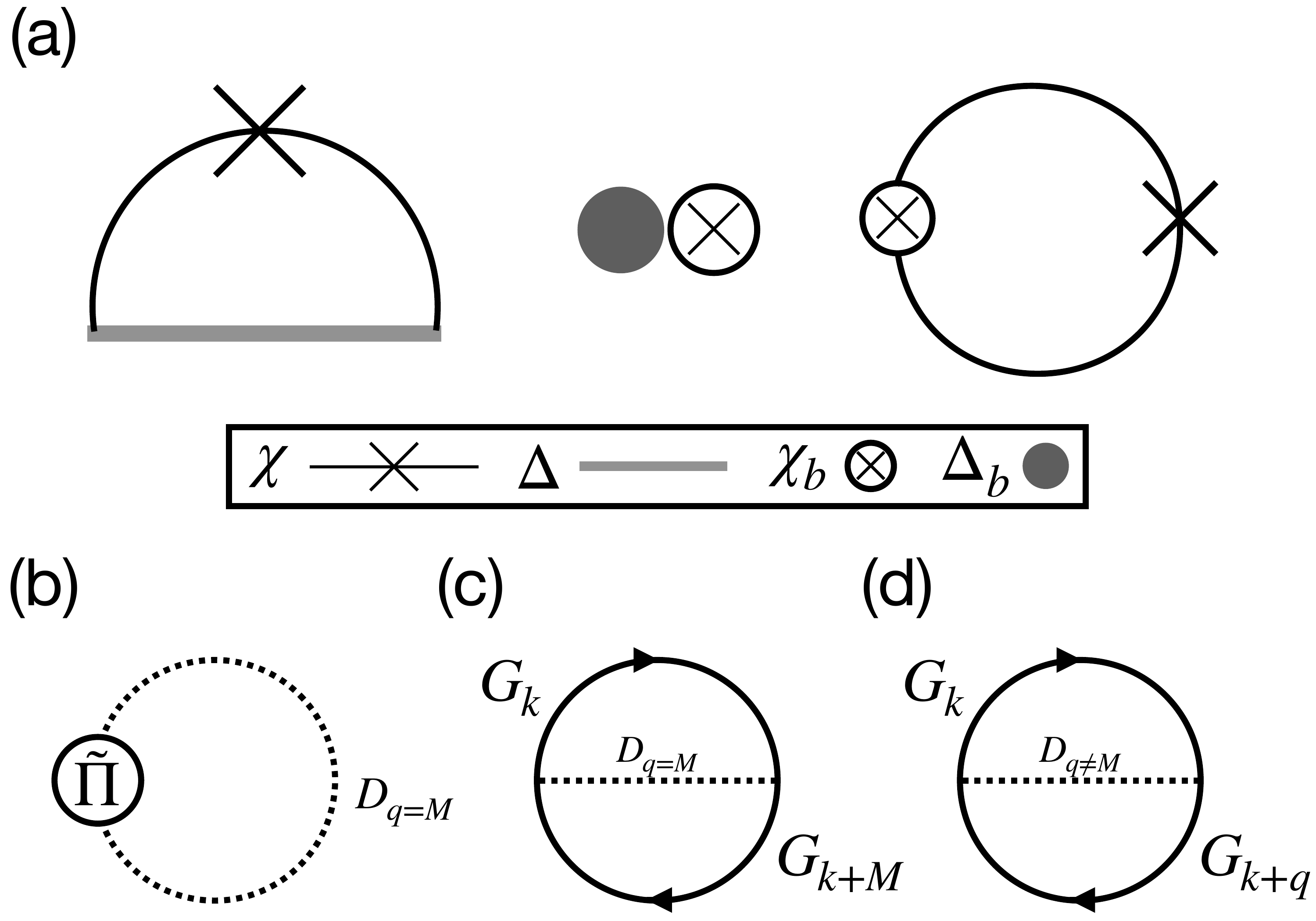}
\caption{Feynman diagrams obtained according to the interacting Hamiltonian, Eq.~(\ref{Eq:new-Hb}).  (a) The lowest order diagrams in the linked-cluster expansion. (b-d) Next-to-leading-order bubble diagrams.}
\label{Fig:Fey}
\end{figure}
\begin{equation}
\D^{ij}_{(\bfk)}=\int \dr \drp \psi^{*}_{i(\bfk)}(\bfr)\D\({\bfr,\bfr'}\)\psi_{j(\bfk)}(\bfr').
\end{equation}
In the following, we will simplify the subscription $(\bfk)~\rightarrow~\bfk$ for density and potential without raising any confusion. On the other hand, for the phonon sector we attribute the phase transition to a single phonon mode $\nu_0$ such that the phonon distortion potential Eq.~(\ref{Eq:Db}) can be reduced to the subspace:
\begin{equation}
\hat{\D}_b=\D^{*}_b\hat{b}_{\bf M \nu_0}+\D_b\hat{b}^\dagger_{\bf M \nu_0}.
\end{equation}
\\
\indent
In practical implementation, we do not directly use Eq.~(\ref{Eq:main-Ha}-\ref{Eq:main-Hc}). Since the DFPT calculation will always provide a screened phonon frequency as discussed in Sec.~\ref{subsect:ph_self}, we extract the screening part from $\hat{H}_I$ and define the frequency adopted in the unperturbed Hamiltonian as:
\begin{equation}
    {\omega^{\rm DFPT}_{\bfq\nu}}=\sqrt{{\omega^{\rm bare}_{\bfq\nu}}^2+2X_{\bfq\nu}(T=0~{\rm K})}
\end{equation}
which becomes imaginary when the phonon is soft. Besides, for the ph-ph anharmonic interactions, we take into account their contributions only in phonon frequency but neglect other scattering effects. As a result, we can write down an effective Hamiltonian for the perturbation theory:
\begin{subequations}
\begin{align}
&\hat{H}_0=
\sum_k \xi^n_{\bfk}\hat{c}_{n\bfk}^\dagger\hat{c}_{n\bfk}
+\sum_{\bfq,\nu}\w^{\rm DFPT}_{\bfq\nu}\hat{b}_{\bfq\nu}^\dagger\hat{b}_{\bfq\nu}
+\tilde{\D}
\label{Eq:new-Ha}
\\
&\hat{H}_I=
\sum_{\bfk\bfq,mn}
g^{\rm DFPT}_{nm\nu}(\bfk,\bfq)
\hat{c}_{n\bfk+\bfq}^\dagger\hat{c}_{m\bfk}
(\hat{b}_{-\bfq\nu}^\dagger+\hat{b}_{\bfq\nu})+c.c.
\nn\\
&~~~~+\sum_{\bfq,\nu\nu'}
\tilde{\Pi}_{\bfq,\nu\nu'}(T)
\hat{b}_{\bfq\nu}^\dagger\hat{b}_{\bfq\nu'}
-\tilde{\D}+({\rm Coulomb})
\label{Eq:new-Hb}
\end{align}
\end{subequations}
where (Coulomb) denotes $\hat{U}^{\rm ee}-\hat{V}_{\rm XC}-\hat{\D}_{\rm XC}$ in Eq.~(\ref{Eq:H_pert1}), and we use the short hands:
\begin{subequations}
\begin{align}
   & \tilde{\D}=\sum_{nm \bfk}\D^{nm}_{\bfk}\hat{c}_{n\bfk+{\bf M}}^\dagger\hat{c}_{m\bfk}
    +\D^{*}_b\hat{b}_{\bf M \nu_0}+c.c.
\\
&\tilde{\Pi}_{\bfq,\nu\nu'}(T)=
\frac{\Omega_{\bfq\v}\Sigma_{\bfq,\nu\nu'}(T)
-\d_{\nu\nu'}X_{\bfq\nu}(T=0~{\rm K})}{\w_{\bfq\nu}^{\rm DFPT}}\label{Eq:Pi_tilde}
\end{align}
\end{subequations}
where $g^{\rm DFPT}$ is the e-ph coupling constant we obtain from first-principle calculation: 
\begin{equation}
    g^{\rm DFPT}_{mn\nu}(\bfk,\bfq)=\(\frac{\hbar}{2\w_{\bfq\v}^{\rm DFPT}}\)^{1/2} \<\psi_{m\bfk+\bfq}|\partial_{\bfq \nu} V|\psi_{n\bfk}\>.
\end{equation}
The self-energy $\Sigma(T)$ is required to be computed by Eq.~(\ref{Eq:Sigma_ph-ph}), and the energy correction Eq.~(\ref{Eq:Pi_tilde}) is chosen for the dressed phonon in this effective Hamiltonian matching the finite-temperature phonon such that the frequency satisfying the self-consistent equation:
\begin{equation}
\label{Eq:w2-w2}
    \Omega^2_{\bfq\v}={\w^{\rm DFPT}_{ \bfq\v}}^2+2X_{\bfq\v}(T)\Big|^{T}_{T=0}+2\Omega_{\bfq\v}\Sigma_{\bfq\v}(T)
\end{equation}
similar to Eq.~(\ref{Eq:w_scp}) in the diagonal approximation (see Appendix~\ref{Append:self_ph}).
\\
\indent
Because the $\Omega_{\bfq\v}(T)$ is the physical pole of the phonon Green's function and corresponds to the frequency measured in experiments, in the rest discussion, we will simply refer "frequency" for this temperature-depending quantity without further specification, and replace $\w^{\rm DFPT}_{\bfq\v}$ by $\Omega_{\bfq\v}(T)$ when computing the thermal average. Note that this substitution also changes the e-ph coupling constant due to the dependence on the phonon frequency via the relation:
\begin{equation}
\label{Eq:gt}
g_{nm\nu}(\bfk,\bfq;T)=g^{\rm DFPT}_{nm\nu}(\bfk,\bfq)\times \sqrt{\frac{\w^{\rm DFPT}_{\bfq\nu}}{\Omega_{\bfq\nu}(T)}}
\end{equation}
where we neglect the temperature effect on the deformation potential and normal mode eigenvector. This replacement of thermal quantities is already adopted in materials with anharmonic dynamics to study the carrier mobility \cite{zhou2018electron}.
\\
\indent
In the following, we apply Eq.~(\ref{Eq:new-Ha}-\ref{Eq:new-Hb}) to derive the formula to study EI/CDW by computing diagrams shown in Fig.~\ref{Fig:Fey}. We temporally focus on the diagram involving only e-ph coupling during the discussion, while the e-e Coulomb interaction will be included in the final result. We divide Fig.~\ref{Fig:Fey} into two categories and investigate their contribution separately. Fig.~\ref{Fig:Fey}(a) presents the lowest order diagrams consisting of density and potentials, while the rest are the self-energy part and bubble diagrams, which will provide the mass renormalization correction. 
\subsection{Gap equation}
\vspace{-10pt}
We first focus on Fig.~\ref{Fig:Fey}(a) and derive the basic structure of the gap equation. The mathematical representation for the diagram is straightforward:
\begin{eqnarray}
    &&U^{\rm (a)}=\sum_{cv\bfk}
    g_{cv\nu_0}(\bfk,{\bf M})\chi^{cv}_{\bfk}
    \(\chi_{b,{\bf M}\nu_0}+\chi^*_{b,{\bf M}\nu_0}\)\nn\\
    &&~~~~~~~~~~~-\D^{cv}_{\bfk} \chi^{cv}_{\bfk}-
    \D_b^*\chi_{b,{\bf M}\nu_0}+c.c.
\end{eqnarray}
where we have used the property, $\hat{b}_{\bf M}=\hat{b}_{-{\bf M}}$. Applying the minimum condition in Eq.~(\ref{Eq:Omega-Omegaks}) and taking the derivative respective to the NL densities, by keeping $\chi_{b,{\bf M}} (\chi_b)$ and $\chi_{b,{\bf M}}^* (\chi^*_b)$ as independent variables, we can obtain:
\begin{subequations}
\begin{align}
&\D_b=\sum_{cv,\bfk} g_{cv\nu_0}(\bfk,{\bf M})\chi^{cv}_{\bfk}+
g^*_{cv\nu_0}(\bfk,{\bf M}){\chi^{cv}_{\bfk}}^*
\label{Eq:gap-eq-Db}\\
&\D^{cv}_{\bfk}=g_{cv\nu_0}(\bfk,{\bf M})
\(\chi_{b,{\bf M}\nu_0}+\chi^*_{b,{\bf M}\nu_0}\)
.
\label{Eq:gap-eq-Dk}
\end{align}
\end{subequations}
These two equations are the same gap equations derived from the mean-field approach \cite{phan2013exciton}, and we will refer them as "MF gap equation" for the rest discussion. They can be solved consistently using the solution of Eq.~(\ref{Eq:new-Ha}) and the definition in Eq.~(\ref{Eq:chi_b}) and Eq.~(\ref{Eq:NL-chi_b}). To present the detail, we carry out an example of an analytically solvable model in a one-dimensional system and left the discussion in Appendix~\ref{Append:MF}.
\subsection{Mass renormalization in $\D_{b}$}
\vspace{-10pt}
The mass renormalization effect can be separated into two parts corresponding to the gap function $\D_{b}$ and $\D_{\bfk}$, respectively, and here we focus on the $\D_{b}$. We isolate the $\bf q=M$ part (Fig.~\ref{Fig:Fey}(c)) from the bubble diagram (Fig.~\ref{Fig:Fey}(d)) and combine it with the contribution from the effective self-energy term. To compute Fig.~\ref{Fig:Fey}(c), we note that due to the $\hat{\D}$ potential term, the phonon operator $\hat{b}^{(\dagger)}_{\bf M}$ in the unperturbed Hamiltonian, Eq.~(\ref{Eq:new-Ha}), acquires a non-zero thermal average such that its thermal Green's function becomes (see Appendix~\ref{Append:ph_G}):
\begin{equation}
D_\nu(\bfq,i\w_n)=
\frac{-2\Omega_{\bfq\nu}}{\w^2_n+\Omega^2_{\bfq\nu}} 
-\d_{\w_n,0}\d_{\bfq,{\bf M}}\d_{\nu,\nu_0}\frac{\b (\D_b^*+\D_b)^2}{\Omega_{\bfq\nu}^2}.
\end{equation}
The Green's function now takes an additional static ($\w_n=0$) term for the $\nu_0$-phonon with $\bfq={\bf M}$ which is quadratically proportional to the potential $\D_b$. 
Therefore, the $\D_b$-relevant part of Fig.~\ref{Fig:Fey}(b,c) becomes:
\begin{eqnarray}
\label{Eq:Ubc}
U^{\rm (b+c)}&=&
\frac{(\D_b^*+\D_b)^2}{\Omega_{{\bf M}\nu_0}^2}
\Pi_{{\bf M}\nu_0}(T)
\nn\\
&+&
\frac{|\D_b|^2}{\Omega^2_{{\bf M}\nu_0}}
\Bigg(
\Sigma_{\bfq,\nu\nu'}(T)
-\Pi_{\bf M\nu_0}(T=0~{\rm K})\Bigg)
\end{eqnarray}
where $\Pi_{\bf M\nu_0}$ is defined as Eq.~(\ref{Eq:ph_self_Pi}) with the e-ph coupling replaced by Eq.~(\ref{Eq:gt}). Using the relation Eq.~(\ref{Eq:b_Db}) and taking the derivative respective to $\chi^*_b$, we can obtain:
\begin{align}
\label{Eq:Fbc_1}
\frac{\partial U^{\rm (b+c)}}{\partial \chi^*_b}
&=\frac{-2\D_b}{\Omega^2_{{\bf M}\nu_0}}
\bigg(2\Omega_{{\bf M}\nu_0}
\Sigma_{{\bf M} \nu_0}(T)+
2X_{{\bf M }\nu_0}(T)\Big|^T_{T=0}
\bigg)\nn\\
&\equiv -\mathcal{Z}_b\D_b
\end{align}
The mass renormalization for the phonon displacement potential can thus be obtained by adding Eq.~(\ref{Eq:Fbc_1}) to the right hand side of Eq.~(\ref{Eq:gap-eq-Db}), such that the gap equation is modified by a overall ratio:
\begin{equation}
\label{Eq:Db_Z} 
\D_b \equiv
\lambda_b 
\[\sum_{cv,\bfk} g_{cv\nu_0}(\bfk,{\bf M})\chi^{cv}_{\bfk}+
g^*_{cv\nu_0}(\bfk,{\bf M}){\chi^{cv}_{\bfk}}^*\]
\end{equation}
where $\lambda_b =1/(1+\mathcal{Z}_b)$. Since the factor $\mathcal{Z}_b$ is always positive, the mass renormalization will reduce the displacement potential and thus decrease the critical temperature for exciton condensation.
\subsection{Mass renormalization in $\D_{\bfk}$ \label{sect:Ze}}
\vspace{-10pt}
The mass renormalization on the NL potential $\D_{\bfk}$ can be obtained from Fig.~\ref{Fig:Fey}(d) which is the bubble diagram involving only normal propagators, and here we include all the phonon momentum $\bfq$ and normal modes $\nu$. To compute Fig.~\ref{Fig:Fey}(d), we first note that for the state of Eq.~(\ref{Eq:tphi}), the Green's function can be written as:
\begin{equation}
\hat{G}_{n(\bfk)}(i\w_n)=
\sum_{cv}
\frac{|t^n_{v\bfk}|^2|{v\bfk}\>\<{v\bfk}|+
|t^n_{c\bfk}|^2|{c\bfk+{\bf M}}\>\<{c\bfk+{\bf M}}|
}
{i\w_n-E_{n(\bfk)} }
\end{equation}
which already mixes states of momentum $\bfk$ and $\bfk+M$ following the Hamiltonian Eq.~(\ref{Eq:new-Ha}). Therefore, Fig.~\ref{Fig:Fey}(d) can be expressed as:
\begin{eqnarray}
\label{Eq:Fd}
U^{\rm (d)}&=&\sum_{\bfk\bfq,nm,\nu}
I(E_{m(\bfk+\bfq),E_{n(\bfk)}},\Omega_{\bfq\nu})\nn\\
&\times&
\bigg[\sum_{v_iv_j}|t^n_{v_i\bfk}|^2|t^m_{v_j\bfk+\bfq}|^2|g_{v_jv_i\nu}(\bfk,\bfq)|^2\nn\\
&&+\sum_{c_ic_j}
|t^n_{c_i\bfk}|^2|t^m_{c_j\bfk+\bfq}|^2|g_{c_jc_i\nu}(\bfk+{\bf M},\bfq)|^2
\bigg]
\nn\\
\end{eqnarray}
where we define
\begin{equation}
I(E_1,E_2,\w)=\sum_{\w_n\w_m}\frac{1}{i\w_n-E_1}\frac{1}{i\w_m-E_2}
\frac{-2\w}{(\w_n-\w_m)^2+\w^2}
\end{equation}
which is the common expression appearing in the bubble diagram including two fermions and one boson propagator. However, in a general system with bands of more than two, there is no analytical expression of coefficient $t^n_{v\bfk}$'s in terms of coupling potential $\D_{\bfk}$. As a result, we take the small $\D_{\bfk}$ approximation and use the perturbation theory to expand the coefficients. In the linear order, since the eigenvalue is the same as  the diagonal entries, we can use the original band index to denote the new states such that:
\begin{eqnarray}
\label{Eq:first-pert}
&&E_{n=v_i(\bfk)}=\xi^{v_i}_{\bfk}~;~
t^{n=v_i}_{v_j\bfk}=\d_{v_i,v_j}~;~
t^{n=v_i}_{c_j\bfk}=\frac{\D^{v_ic_j}_{\bfk}}{\xi^{c_j}_{\bfk+{\bf M}}-\xi^{v_i}_{\bfk}}
\nn\\
&&E_{n=c_i(\bfk)}=\xi^{c_i}_{\bfk+{\bf M}}~;~
t^{n=c_i}_{c_j\bfk}=\d_{c_i,c_j}~;~
t^{n=c_i}_{v_j\bfk}=\frac{\D^{c_iv_j}_{\bfk}}{\xi^{v_j}_{\bfk}-\xi^{c_i}_{\bfk+{\bf M}}}\nn\\
\end{eqnarray}
However, the linear order is insufficient to express the eigenvector terms in $U^{\rm (d)}$ since the terms always appear as an absolute square, $|t^n_{c,v}|^2$. We compute the perturbation expansion to the second order for the explicit $\D^{cv}_{\bfk}$-dependence and take the derivative on Eq.~(\ref{Eq:Fd}). The algebraic detail is left to the Appendix~\ref{Append:dUdchi}, and we quote the result here. The derivative of the bubble diagram respective to NL-densities can be written down as:
\begin{align}
\label{Eq:Fddchi}
&\frac{\partial U^{\rm (d)}}{\partial \chi_{\bfk}^{cv*}}
=\frac{\D^{cv}_{\bfk}}{(\xi^{c}_{\bfk+{\bf M}}-\xi^{v}_{\bfk})
\(f_{\bfk+{\bf M},c}-f_{\bfk,v}\)}\nn\\
&~~~~~\times\sum_{\bfq,\nu}\bigg[\sum_{v_2}I(\xi^{v_2}_{\bfk+\bfq},\xi^{v}_{\bfk},\w_{\bfq\nu}) |g_{v_2v\nu}(\bfk,\bfq)|^2\nn\\
&~~~~~~~~~~~~~+\sum_{c_2}
I(\xi^{c_2}_{\bfk+{\bf M}+\bfq},\xi^{c}_{\bfk+{\bf M}},\w_{\bfq\nu}) |g_{c_2c\nu}(\bfk+{\bf M},\bfq)|^2\bigg]\nn\\
&~~~~~~~~\equiv -\mathcal{Z}^{cv}_{\bfk}\D^{cv}_{\bfk}.
\end{align}
By adding Eq.~(\ref{Eq:Fddchi}) to the right-hand side of Eq.~(\ref{Eq:gap-eq-Dk}), the gap equation is then modified by an overall ratio from the mass renormalization term:
\begin{figure*}[t]
\centering
\includegraphics[scale=0.265]{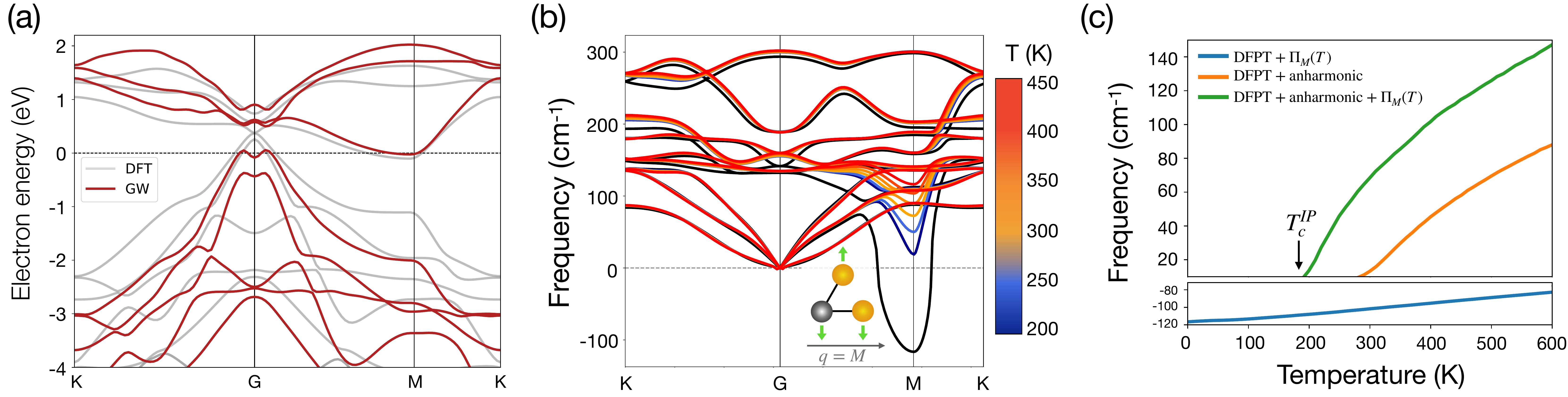}
\caption{ (a) Band structure of monolayer $\rm TiSe_2$ in disordered phase computed in DFT (grey) and the GW approximation (deep red) with spin-orbit coupling included. The electron pocket at the M point and the hole pocket at the $\Gamma$ point shrink after the GW correction included. (b) The temperature dependence of the SCP phonon dispersion is encoded by the color map. The black line is computed by DFPT in the harmonic approximation, for which the appearance of the soft mode at the M point indicates the instability of the crystal structure against the formation of $2\times 2$ CDW at zero temperature. Inset: The normal vibration mode of the soft mode at the M point. (c) The soft mode frequency as a function of temperature with the correction from the screening effect and ph-ph anharmonic interaction. In the independent particle (IP) approximation, by neglecting the exciton condensation, we obtain a critical temperature $T^{\rm IP}_c=190~$K.}
\label{Fig:bands}
\end{figure*}
\begin{equation}
    \label{Eq:Dk_Z}
    \D^{cv}_{\bfk}\equiv \lambda^{cv}_{\bfk}
     g_{cv\nu_0}(\bfk,{\bf M})
\(\chi_{b,{\bf M}\nu_0}+\chi^*_{b,{\bf M}\nu_0}\)
\end{equation}
where $\lambda^{cv}_{\bfk}=1/(1+\mathcal{Z}^{cv}_{\bfk})$.
Due to the occupation difference in the denominator, $\mathcal{Z}_{\bfk}$ diverges and suppresses the potential $\D_{\bfk}$ for pairing between electron-electron($e$-$e$) and hole-hole($h$-$h$), such that the remaining configuration can only contain one electron and one hole ($e$-$h$ pairing).
\\
\indent
In conclusion, Eq.~(\ref{Eq:Db_Z}) and Eq.~(\ref{Eq:Dk_Z}) determine the gap which characterizes the excitonic insulator triggered by the e-ph coupling in a distorted lattice structure. The results contain the crucial mass renormalization factors for the electronic and atomic sectors, which are missed in the conventional MF approach. This derivation starts from first principles, requires no empirical parameters, and can be applied to study natural systems. 
\section{Exciton condensation in monolayer $\bf TiSe_2$ \label{Sect:ex_TiSe2}}
\vspace{-10pt}
In this section, we present the results for the EI/CDW phase in monolayer  $\rm TiSe_2$.
Although the EEL experiment resolves the debate on bulk $\rm TiSe_2$ \cite{kogar2017signatures} as discussed in the introduction, a similar measurement for its monolayer crystal is still missing in the literature. On the other hand, current $ab~initio$ studies can not cooperate the electron and phonon in a unified approach \cite{singh2017stable,guster20182,zhou2020anharmonicity} and the e-h pairing mediated by phonon coupling has never been elucidated yet. Therefore, to reveal the underlying mechanism of the unconventional CDW in monolayer $\rm TiSe_2$, we apply the formalism developed in the last section to carry out the numerical simulation for the exciton condensation and discuss all the factors affecting the critical temperature from first principles.
\subsection{electronic structure of monolayer $\rm TiSe_2$ in disordered phase}
\vspace{-10pt}
The first-principles electronic structure of monolayer $\rm TiSe_2$ can be found in the literature \cite{singh2017stable}, and here we present our calculation as a starting point. We perform the DFT calculation on the relaxed hexagonal structure using the {\sc QUANTUM ESPRESSO} package \cite{giannozzi2009quantum}, with the Perdew–Burke–Ernzerhof exchange-correlation functional and fully relativistic norm-conserving pseudopotentials generated with Pseudo Dojo \cite{troullier1991efficient,perdew1996generalized,van2018pseudodojo}. Going beyond DFT, we use the VASP code to include the GW correction in an additional calculation \cite{kresse1996efficient}. All the computational details are summarized in Appendix~\ref{Append:numerical}. 
\\
\indent
In Fig.~\ref{Fig:bands}(a), we present the computed band structure where we match two results by their Fermi level at 0 eV and connect isolated data points by the Wannier interpolation method using the interface between {\sc VASP} and {\sc wannier90} \cite{mostofi2014updated}.
Both DFT and GW calculation shows that monolayer $\rm TiSe_2$ is a semi-metal with a hole pocket near the $\Gamma$ point and electron pocket in the M point while the Fermi surface is much smaller in the GW bands. By resolving the orbital in the band structure \cite{singh2017stable}, we found that, with respect to the Fermi level, bands containing the Se {\it p-}orbitals tend to be suppressed to lower energy while the bands with the Ti {\it d-}orbitals can be promoted to higher energy in the GW correction. Near the $\Gamma$ point, two kinds of orbital are entangled, so the bands have Mexican hat-like dispersions. On the other hand, away from the $\Gamma$ point, the overall vertical stretch between occupied and unoccupied bands is about 1 eV.
\subsection{Temperature depending phonon dispersion}
\begin{figure}[t]
\includegraphics[scale=0.2]{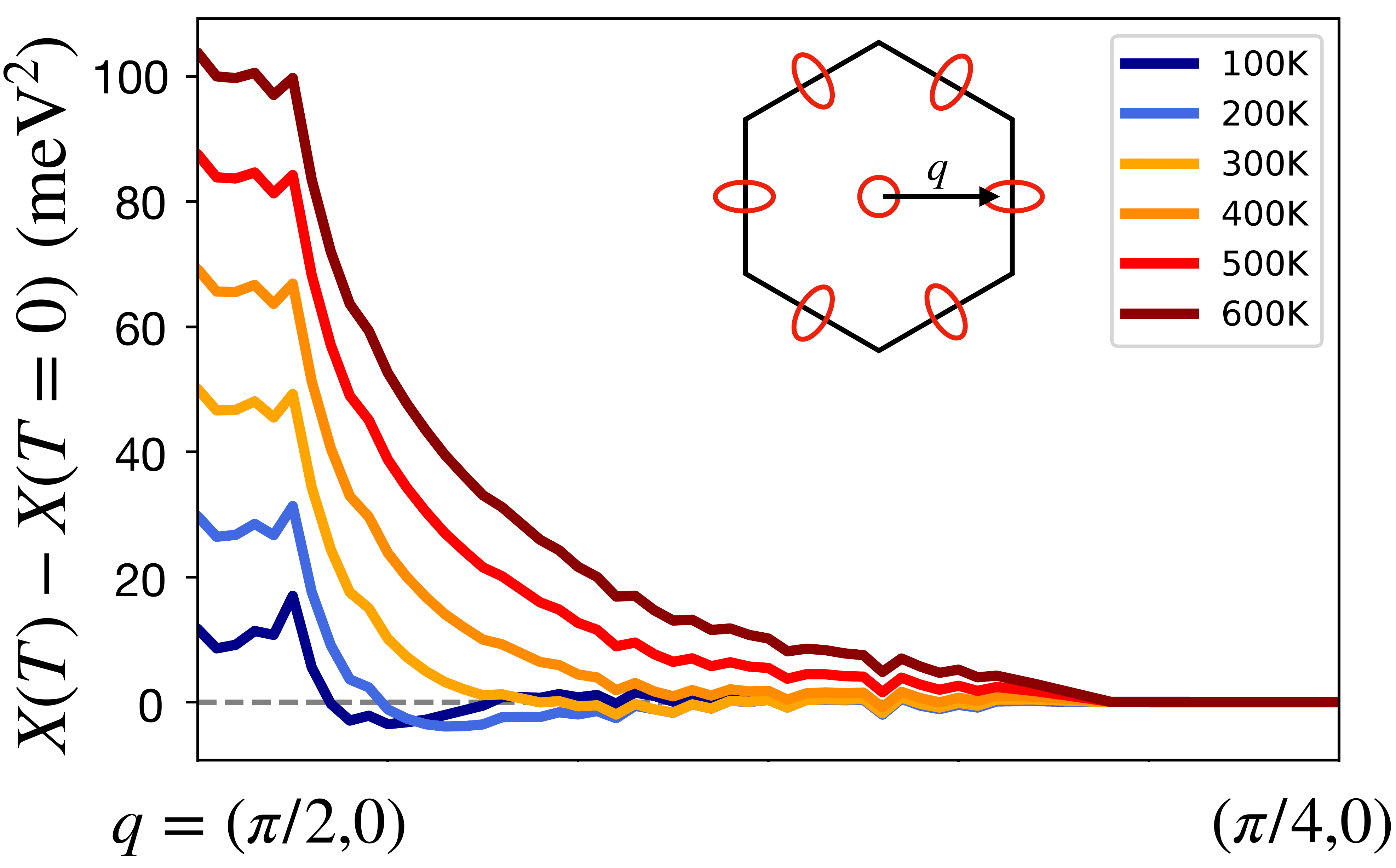}
\caption{Coulomb screened self-energy correction at finite temperature, Eq.~(\ref{Eq:wPi}), measured with respect to the zero temperature value. Although the Fermi surface nesting is not significant (circle vs. oval), remarkable temperature dependence can be found for phonon near ${\bf M}=(\pi/2,0)$ connecting the electron packet and hole pocket.}
\label{Fig:Pi}
\end{figure}

\vspace{-10pt}
In bulk $\rm TiSe_2$ a soft phonon with imaginary frequency has been observed and computed with wave vector $\bfq=L=(\frac{\pi}{2},\frac{\pi}{2},\frac{\pi}{2})$ which corresponds to the formation of $2\times 2\times 2$ CDW in three-dimensional crystal \cite{holt2001x,duong2015ab,subedi2022trigonal}. The soft phonon can be "hardened" by increasing the environment temperature.  However,  in first-principles studies where
occupation broadening was used for charge density to simulate the finite temperature effect, extremely large broadening is required for dynamical stability and unrealistic high transition temperature was obtained \cite{duong2015ab,singh2017stable} for both the bulk and monolayer structures. The failure of this approach indicates a strong anharmonicity in $\rm TiSe_2$ beyond the harmonic approximation, which is highlighted in a recent study \cite{zhou2020anharmonicity}. 
\\
\indent
Here we apply the method introduced in Sect.~\ref{subsect:ph_self} to compute the self-consistent anharmonic phonon frequency and study the phonon softening in monolayer $\rm TiSe_2$. In the practical procedure, we carry out the DFPT calculation using {\sc QUANTUM ESPRESSO} to obtain the dynamical matrix and compute the e-ph matrix element with {\sc PERTURBO} code \cite{zhou2021perturbo} which is then used in Eq.~(\ref{Eq:ph_self_Pi}) to compute the screened self-energy. On the other hand, to compute the phonon anharmonicity, we first use the {\it ab initio} molecular dynamics method implemented in {\sc VASP} to generate the interatomic force and then compute the anharmonic phonon frequency by the {\sc ALAMODE} code \cite{tadano2018first}. 
\\
\indent
We present the temperature-dependent phonon dispersion in Fig.~\ref{Fig:bands}(b), where we plot the imaginary frequency in the minus axis. The black line showing a robust soft mode at the M point is the direct result of DFPT, which is consistent with the formation of $2\times 2$ CDW of monolayer $\rm TiSe_2$. In the inset, we plot the corresponding normal mode for the soft phonon, which has atomic displacement perpendicular to the wave vector.
In the hexagonal Brillouin zone (BZ), there are three inequivalent M points, and 
the condensation of these three normal modes can build up the lattice distortion observed in the CDW phase. Beyond the harmonic approximation, we can see that, by disregarding some temperature-independent correction, the soft phonon is the only mode sensitive to the anharmonic effect. Overall, the properties presented in the above results validate the single-mode assumption applied in Sect.~\ref{Sect:ex-cond22}.
\\
\indent
In Fig.~\ref{Fig:bands}(c), we analyze the soft-phonon at the M point. We present the phonon frequency as a function of temperature. We consider three different cases: that considers the correction only from the screening (Eq.~(\ref{Eq:ph_self_Pi})), that considers the correction only from the anharmonic effect (Eq.~(\ref{Eq:Sigma_ph-ph})), and that considers 
both the two effects. First, we note that, at zero-temperature, the harmonic DFPT gives us a phonon frequency $\w^{\rm DFPT}_{\bf M}=116i~{\rm cm^{-1}}$. When considering the temperature dependence in the Coulomb screened self-energy, the frequency varies by $30i~{\rm cm^{-1}}$ from 0~K to 600~K, which indicates that the phonon will always be soft in the harmonic approximation. On the other hand, the pure anharmonic effect can raise the phonon frequency dramatically, forcing the phonon mode to become "hard" above $T=270$~K. This critical temperature can even be reduced to $T=190$~K when we combine the contribution from the two effects and solve the phonon frequency self-consistently. 
In Fig.~\ref{Fig:Pi}, we present the self-energy change with respect to the zero-temperature value for phonon momentum $\bfq$ along the high symmetry line M-$\Gamma$ under different temperatures. We note that the most significant temperature dependence does not occur for the momentum $\bfq={\bf M}$ since the Fermi surface is not ideally "nesting." Instead, it is approximately a constant for $\bfq$ near the M point within a specific range and gradually decreases outside.
\subsection{Gap equation}
\begin{figure}[t]
\centering
\includegraphics[scale=0.14]{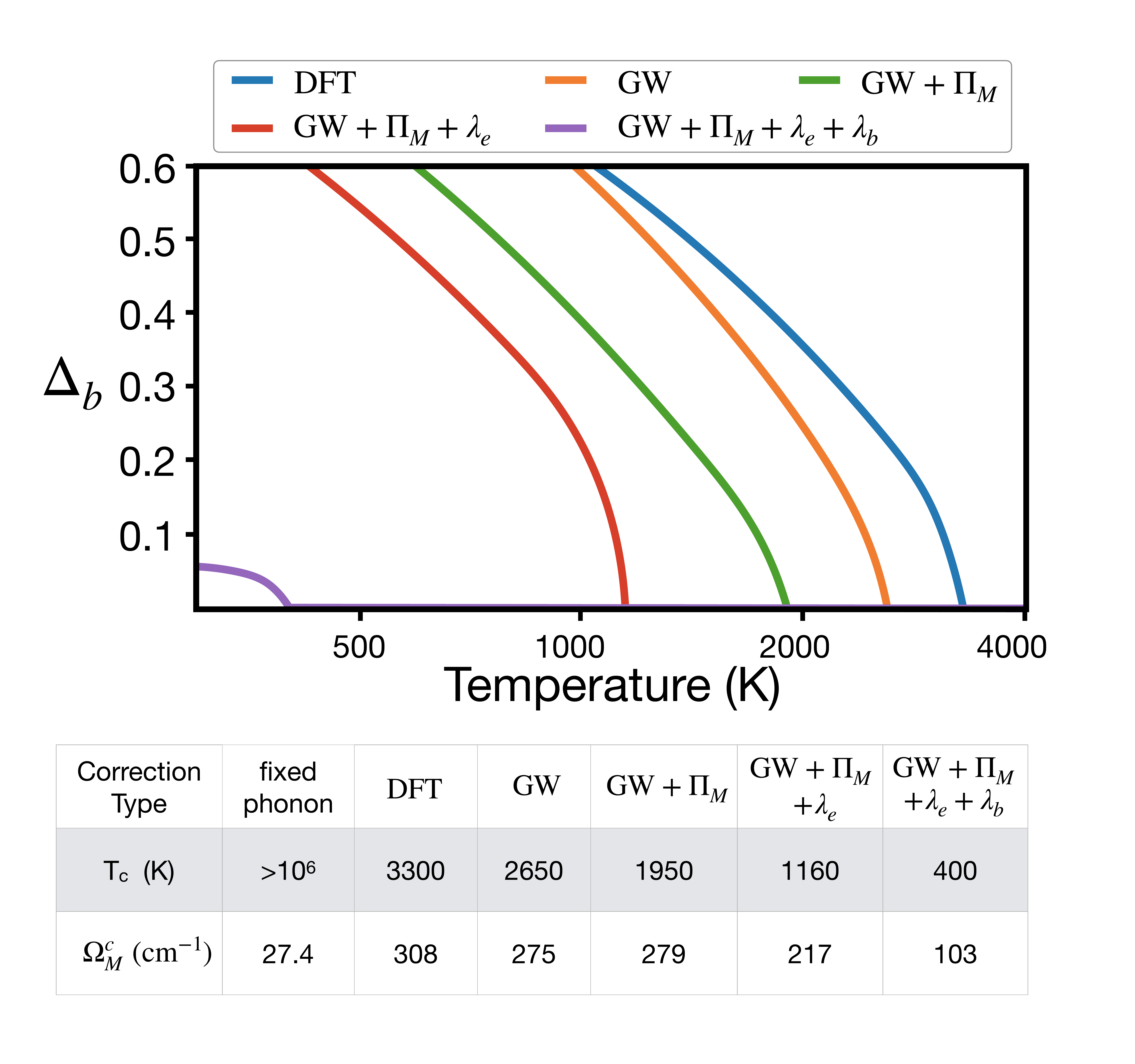}
\caption{Temperature depending phonon displacement potential computed under different corrections. The inclusion of the phonon mass renormalization term suppresses the potential in the whole temperature range, while other corrections provide a rigid shift. Table: Summaries of critical temperature and corresponding phonon frequency.}
\label{Fig:Db_compare}
\end{figure}

\begin{figure}
\centering
\includegraphics[scale=0.16]{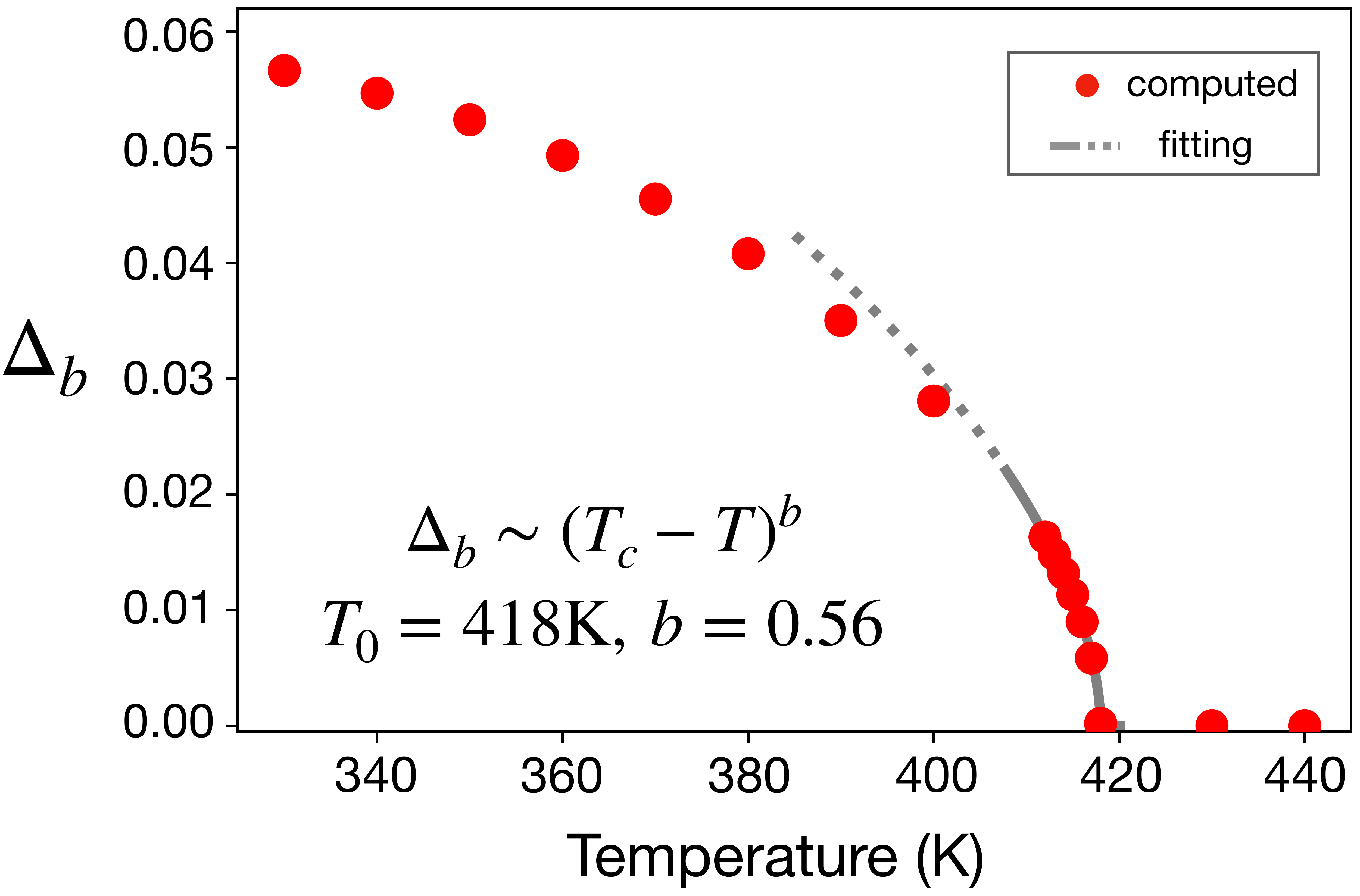}
\caption{EI/CDW potential calculated from first-principles. The computed critical temperature is $T_c=418~$K while the experimental measure is $T^{\rm exp.}_c=220~$K. The fitted critical component $b=0.56$ is in great agreement with the square root trend in the experiment \cite{fang2017x}.}
\label{Fig:Db_all}
\end{figure}

\vspace{-10pt}
Based on the discussion in the last section, the EI/CDW lattice distortion is attributed to the combined effect of the three inequivalent M phonons. This so-called "triple-$\bfq$" effect requires us to extend the "single-$\bfq$" formalism presented in Sect.~{\ref{Sect:ex-cond22}}. More explicitly, we write down the new wave function by introducing the pairing direction in the Eq.~(\ref{Eq:tphi}):
\begin{equation}
    \phi^{n}_{(\bfk)}=
    \sum_v t^{n}_{v\bfk} \psi_{v\bfk}+
    \sum_{c,I} t^{n}_{c\bfk,I} \psi_{c\bfk+{\bf M}_I}
\end{equation}
where $I=1,2,3$ is the index for the three {\bf M} directions. In a simplified two band model, the mixing coefficients can be obtained by solving the triple-$\bfq$ version of the Hamiltonian Eq.~(\ref{Eq:new-Ha}):
\begin{equation}
\label{Eq:gap-matrix}
\begin{pmatrix} 
\xi ^{ c }_{ \bfk+{\bf M}_{ 1 } } & 0 & 0 & \Delta ^{ cv}_{ \bfk,1 }  \\ 
0 & \xi ^{ c }_{ \bfk+{\bf M}_{ 2 } } & 0 & \Delta ^{ cv }_{ \bfk,2 }   \\ 
0 & 0 & \xi ^{ c }_{ \bfk+{\bf M}_{ 3 } } &  \Delta^{cv}_{\bfk,3} \\ 
\Delta ^{ v c}_{\bfk,1} & \Delta ^{ vc }_{ \bfk,2 } &\Delta ^{ vc}_{ \bfk,3 }  & \xi ^{ v }_{ \bfk }  
   \end{pmatrix}
   \begin{pmatrix} 
  t^{n}_{c\bfk,1}\\ t^{n}_{c\bfk,2}\\ t^{n}_{c\bfk,3}\\ t^{n}_{v\bfk}
   \end{pmatrix}
   =E_{n(\bfk)}\begin{pmatrix} 
   t^{n}_{c\bfk,1}\\ t^{n}_{c\bfk,2}\\ t^{n}_{c\bfk,3}\\ t^{n}_{v\bfk}
   \end{pmatrix},
\end{equation}
where we neglect the pairing between states of $\bfk+{\bf M}_I$ and $\bfk+{\bf M}_J$ for which the momentum transfer exceeds the first BZ. It can be shown that the effect of triple-$\bfq$ will decouple from each other in the limit of $\D_{\bfk} \rightarrow 0$ and reduce to the single-$\bfq$ scenario in the linear order as Eq.~(\ref{Eq:first-pert}). This property reflects that the inclusion of the triple-$\bfq$ effect only changes the finite $\D_{\bfk}$ behavior but maintains the critical temperature, $T_c$ where $\D_{\bfk}=0$. To correctly describe the behavior below $T_c$, in the implementation, we will construct and solve the gap equation with the triple-$\bfq$ effect included.
\\
\indent
To comprehensively understand the role of each component in the gap equation, we first note that the temperature dependence in the EI/CDW gap equations lies in: electronic occupation, phonon frequency, electron-phonon coupling, and mass renormalization coefficients, and each of the factors is crucial to compute the critical temperature. To emphasize the importance, we first apply the MF gap equations, Eq.~(\ref{Eq:gap-eq-Db}) and Eq.~(\ref{Eq:gap-eq-Dk}), as a baseline and use the DFT band structure with the phonon frequency near the experimental critical temperature, $T^{\rm exp.}_c=220~$K. However, the result gives an extreme high $T_c$, since the small frequency, $\Omega_{\bf M}(T_c^{\rm exp.})\approx 30 ~{\rm cm^{-1}}$, can boost the phonon thermal average with small potential $\D_b$ and enhance the coupling constant by Eq.~(\ref{Eq:gt}). These two effects make the critical temperature higher than $10^6$ K.
\\
\indent
To analyze how the phonon frequency affects the critical temperature, we restrict discussions in the MF gap equation and study three different scenarios. First, we restore the temperature dependence in the anharmonic phonon frequency without correction on Coulomb screened self-energy. A great improvement in the critical temperature is obtained (see blue line in Fig.~\ref{Fig:Db_compare}), which gives $T_c\sim 3300$~K with the critical phonon frequency $\Omega^c_{\bf M}=308 ~{\rm cm^{-1}}$. Compared to the fixed-frequency case, a ten-times enhancement on $\Omega_{\bf M}$ can suppress the critical temperature by three orders of magnitude. This modification reflects that in a BCS-like mean-field theory, $T_c \sim \exp(1/U)$, where the pairing potential $U$ is characterized by $|g_{\bf M}^2|/\Omega_{\bf M}$ here \cite{micnas1990superconductivity,yuh2006excitonic}. 
\\
\indent
Next, we add the GW correction to increase the relative energy between occupied and unoccupied states. The result with a modified band structure differs from the calculation with DFT bands only at temperatures higher than 1200~K (see the orange line in Fig.~\ref{Fig:Db_compare}); on the other hand, at temperatures lower than 1200~K, the coupling $\D_{\bfk}^{cv}$ is much larger than 1 eV, which makes the GW correction irrelevant. Last, fixed on the GW band, we use the phonon frequency dressed by the temperature-dependent Coulomb screening, which can suppress the lattice distortion potential in the whole temperature ranges and lower the $T_c$ to 1950 K. We found that although the $T_c$ is significantly reduced, the critical frequency $\Omega^c_{ \bf M}$ is merely changed  (see the table in Fig.~\ref{Fig:Db_compare}). Thus, we conclude that in the mean-field approach, the major temperature-dependent factor in EI/CDW phase transition is dominated by the phonon frequency; disregarding the temperature, as long as the phonon has a frequency lower than $\sim 275~{\rm cm^{-1}}$, the lattice structure becomes unstable and transit into the distorted $2\times 2$ superlattice.
\begin{figure}[t]
\centering
\includegraphics[scale=0.31]{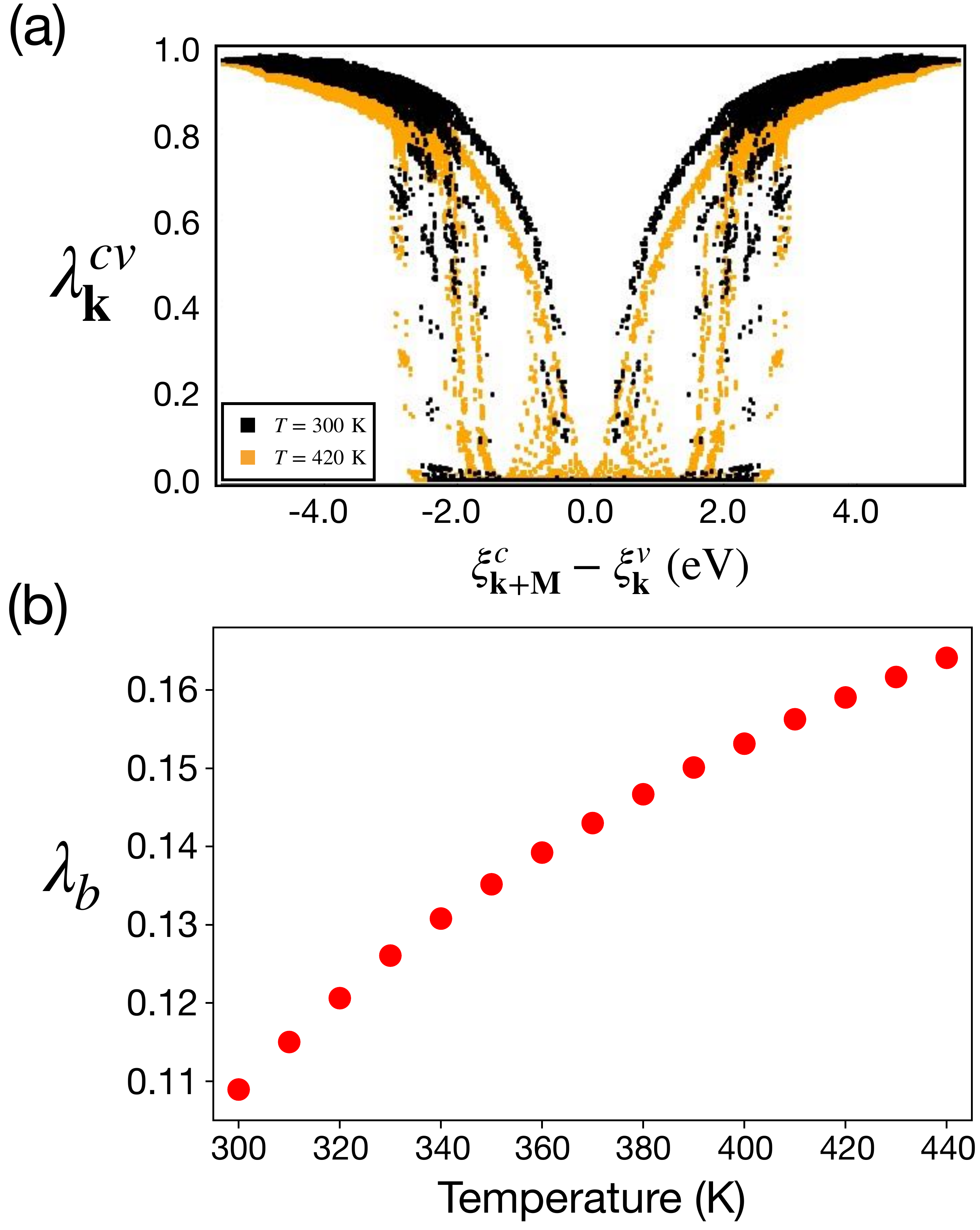}
\caption{Mass renormalization coefficients near the critical temperature.  (a) The exciton condensation mass renormalization coefficients can be categorized into three kinds: those coming from the e-e(h-h) pairing, energy modulation, and pairing involving states near the Fermi surface, respectively. The overall values decrease when temperature rises. (b) Phonon displacement mass renormalization coefficient increases along with temperature.}
\label{Fig:Zs}
\end{figure}
\\
\indent 
The critical frequency $\Omega^c_{\bf M}\sim 275~{\rm cm^{-1}}$ obtained above is extremely high, given the highest phonon mode in the full dispersion is about $300~{\rm cm^{-1}}$ (see Fig.~\ref{Fig:bands}(b)), which indicates the insufficiency of using the MF gap equation. In the following, we discuss the mass renormalization effect by considering Eq.~(\ref{Eq:Db_Z}) and Eq.~(\ref{Eq:Dk_Z}) beyond the MF approach. In Fig.~\ref{Fig:Zs}(a), we present the electron mass renormalization coefficient $\lambda_{\bfk}^{cv}$ as a function of energy transition. The result shows a symmetric pattern with respect to the zero-energy transition since we involve the same number of bands for states with wave vector $\bfk$ and $\bfk+{\bf M}$; for each $v\bfk$ and $c\bfk+{\bf M}$, there is always a corresponding $v\bfk+{\bf M}$ and $c\bfk+2{\bf M}$ which has the opposite energy difference. Overall, $\lambda_{\bfk}^{cv}$ can be categorized into three types. 
The first is the zero values ranging from $-2.0$ eV to $2.0$ eV. They correspond to the mixing between bands below $(h$-$h)$ or above $(e$-$e)$ the Fermi level  as discussed in Sec.~\ref{sect:Ze}. 
The second is the smooth change from 1.0 to 0.0 when $\xi^{c}_{\bfk+{\bf M}}-\xi^{v}_{\bfk}$ approaches zero. This behavior is controlled by the divergence of  $\mathcal{Z}_{\bfk}^{cv}$ due to the energy difference term in the denominator in Eq.~(\ref{Eq:Fddchi}). 
It should be noted that for cases with small $\xi^{c}_{\bfk+{\bf M}}-\xi^{v}_{\bfk}$ the vanishing $\lambda_{\bfk}^{cv}$ will not forbid the pairing as long as occupation difference $f_{{\bf k+M},c}-f_{\bfk,v}$ is finite. Taking Eq.~(\ref{Eq:gap-matrix}) as an example, although the pairing potential $\D_{\bfk}^{cv}$ is weak in this scenario, the relative spacing among diagonal terms $\xi^i_{\bfk}$ is also small such that the resultant eigenvector $t^n_{i\bfk}$ can still create a significant pairing effect. 
The third is the vertical scattered distribution near $\xi^{c}_{\bfk+{\bf M}}-\xi^{v}_{\bfk}\approx \pm 2.0$ eV which is the intermediate type of the previous two. $\lambda_{\bfk}^{cv}$ of this kind comes from the pairing involving states near the Fermi-surface where the pairing configuration can easily change from $e$-$h$ to $e$-$e$ or $h$-$h$ pairing and reduce the $\lambda_{\bfk}^{cv}$ within a small energy variation.
Overall, once the temperature is raised from $T$=300~K to $T$=420~K, as contrasted in Fig.~\ref{Fig:Zs}(a), the $\lambda_{\bfk}^{cv}$ decreases and weakens the pairing potential, which makes it faster to reach the critical temperature in the heating process. 
On the other hand, in Fig.~\ref{Fig:Zs}(b), we present the mass renormalization coefficient from the phonon sector $\lambda_b$ as a function of temperature. It shows that the coefficient increase as temperature rises. The result is in the opposite trend from the $\lambda^{cv}_{\bfk}$'s because the $\mathcal{Z}_b$ factor is inversely proportional to the phonon frequency such that smaller $\Omega_{\bf M}$ can result in a smaller $\lambda_b$ and vice versa.
\\
\indent
Using the renormalization coefficients, we first compute the gap equation with $\lambda_{\bfk}^{cv}$ and then add on the effect from $\lambda_b$ and present the corresponding gap function in Fig.~\ref{Fig:Db_compare}. Each correction can reduce the critical temperature by $\sim 800~$K, respectively. For $\D_b >0.3$, $\lambda_{\bfk}^{cv}$ contribute a rigid shift lowering the $\D_b$ by 0.1 while, below 0.3, $\D_b$ becomes unstable and drop to zero within $\Delta T=$200 K (red line). Therefore, the character of second-order transition becomes more significant with $\lambda_{\bfk}^{cv}$ compared to the mean-field curves \cite{lopes2022excitonic}. On the other hand, besides reducing the critical temperature, $\lambda_b$ even modifies the atom displacement potential within the EI/CDW phase. Without $\lambda_b$, $\D_b$ will blow up exponentially as temperature decreases, but $\lambda_b$ can significantly suppress the $\D_b$ at a lower temperature. So that the $\D_b-T$ curve becomes flat except near the critical temperature (purple line). Now, at the critical temperature, $T_c=400$~K, the corresponding critical frequency $\Omega_{\bf M}^c=103~{\rm cm^{-1}}$ which is comparable to the folded phonon frequency computed at $\Gamma$-point $\Omega_{{\bf M}\rightarrow\Gamma}\sim 75~{\rm cm^{-1}}$ in the $2 \times 2$ CDW phase \cite{singh2017stable}.
\\
\indent
Before closing the discussion, we restore the electron-hole Coulomb attraction into the gap equation where the screened Coulomb matrix elements are extracted using the BSE-subroutine in the {\sc YAMBO} codes \cite{sangalli2019many}. We present the result as one of the main conclusions in this work in Fig.~\ref{Fig:Db_all}. The inclusion of the Coulomb force only raises the critical temperature by 20 K. Without showing the result, we note that with the Coulomb attraction as the only mechanism in the MF approach, the critical temperature is $\sim 600$~K, much lower than the $T_c=1950$ by e-ph coupling. Besides, the Coulomb kernel is restricted to pairing near the Fermi surface, which is thus easier to be restrained by the mass renormalization factor. As a result, we conclude that the underlying mechanism causing the formation of the CDW order in monolayer $\rm TiSe_2$ is the strong electron-phonon coupling mixing the occupied and unoccupied states. At the same time, the screened Coulomb attraction can only provide a minor contribution during the phase transition. Fitting the discrete data near the critical point, we obtain the critical temperature $T_c=418$~K with a critical component $b=0.56$ in the temperature dependence $\D_b\sim (T_c-T)^b$ which is consistent with the experimental observation \cite{fang2017x}.

\section{Conclusion \label{Sect:conclusion}}
\vspace{-10pt}
In this work, we presented a general approach extending the DFT to describe the structural phase transition due to the exciton condensation. In the framework, we introduce the non-local density to encode the spontaneous formation of exciton pairs by the Coulomb attraction and e-ph interaction. For the phonon sector, we include the phonon anharmonicity by the self-consistent phonon theory to harden the soft-phonon appearing in DFPT while bringing in a linear distortion potential to measure the lattice deformation. The exchange-correlation function in this generalized DFT can be computed by the many-body perturbation theory with the assistance of the linked-cluster expansion such that a gap equation to describe the EI/CDW can be derived self-consistently. In practical application, we use the formalism to study the 1T-$\rm TiSe_2$ monolayer crystal. The numerical results favor our method over the mean-field approach by crucial mass renormalization effects in both electron and phonon states. We obtained a second-order phase transition between a semi-metal and a $2\times 2$ CDW caused by e-ph coupling with critical temperature $T_c=420$~K compared to the experimental value of 220~K. 
To our knowledge, this work is the first {\it ab initio} study that analyzes the role of ph-ph, e-ph, and e-e interactions on the same footing for  $\rm TiSe_2$. In conclusion, our work provides a framework for predicting the exciton condensation and studying its relation to lattice instability from first principles. The methodology is general to be applied to other excitonic insulators beyond the example studied in this work. These calculations can guide the future search for candidate material with the excitonic insulating phase.

\begin{acknowledgments}
We are grateful for fruitful discussions with A. S. Mishchenko and R. Masuki. We acknowledge the financial support by Grant-in-Aids for Scientific Research (JSPS KAKENHI) [Grant No. JP19H05825] and “Program for Promoting Researches on the Supercomputer Fugaku” (Project ID: hp220166) from MEXT, Japan.
\end{acknowledgments}

\appendix

\section{GW correction from NL density \label{Append:GW}}
\vspace{-10pt}
Here we use Eq.~(\ref{Eq:H_pert1}) to derive the correction due to the NL XC potential $\hat{\D}_{\rm XC}$ and show the agreement between this approach and the GW-correction. Using the electron operator, we can expand Eq.~(\ref{Eq:KS-NLH}) in an explicit form:
\begin{align}
&\hat{H}=\int d\bfr^3 \Psi^\dagger(\bfr)\left[\frac{-\nabla^2}{2}+v_{\rm KS}(\bfr)-\mu\right]\Psi(\bfr)\nn\\
&+
\int d\bfr^3d{\bfr'}^3\left[\Delta(\bfr,\bfr')\Psi^\dagger(\bfr)\Psi(\bfr')+h.c.\right]
\end{align}
where we have dropped the spin index $\s$. Assuming this Hamiltonian can be diagonalized by orbital wave functions $\phi_{m}(\bfr)$ and operators $\hat{c}_{m}^{(\dagger)}$ such that we can expand the electron operator:
\begin{equation}
\Psi(\bfr)=\sum_{n}\hat{c}_{n}\phi_{n}(\bfr),~{\rm where}~~
\<\hat{c}^\dagger_{m}\hat{c}_{n}\>=\d_{mn}f_n,
\end{equation}
where $n$ is the orbital index, and $\b=1/k_BT$. Thus, we can write down the Schrodinger equation:
\begin{align}
\label{Eq:schrodinger+delta}
E_{n}\phi_{n}(\bfr)&=\left[\frac{-\nabla^2}{2}+v_s(\bfr)-\mu\right]\phi_{n}(\bfr)
\nn\\
&+\int \drp \[\Delta(\bfr,\bfr')+\Delta^*(\bfr',\bfr)\]\phi_{n}(\bfr').
\end{align}
Based on the orbital wave function, we can re-write the charge density:
\begin{align}
n(\bfr)\equiv
\sum_{n}n_{n}(\bfr)=\sum_{n}f_n
|\phi_{n}(\bfr)|^2
\end{align}
and the NL density:
\begin{align}
\label{Eq:GW-chi}
\chi(\bfr,\bfr')&=\sum_{n}\chi_{n}\phi^*_{n}(\bfr)\phi_{n}(\bfr')
\end{align}
where $\chi_{n}$ is now the parameter for NL density in the orbital basis and can be shown to satisfy $\chi_{n}=f_n$. Generally in the GW approximation, the additional NL XC potential is assumed to be diagonal in the basis of DFT orbitals. Therefore, for eigenfunctions satisfying the KS equation:
\begin{equation}
\label{Eq:schrodinger-DFT}
\xi^{n}\bar{\phi}_{n}(\bfr)=\left[\frac{-\nabla^2}{2}+v_{s}(\bfr)-\mu\right]\bar{\phi}_{n}(\bfr),
\end{equation}
it also satisfies Eq.~(\ref{Eq:schrodinger+delta}) such that we have ${\phi}_{n}(\bfr)=\bar{\phi}_{n}(\bfr)$ and:
\begin{align}
\int \dr\drp \bar{\phi}^*_{n}(\bfr) 
\[\Delta(\bfr,\bfr')+\Delta^*(\bfr',\bfr)\]\bar{\phi}_{m}(\bfr')
\equiv\d_{nm} \D_{n}\label{Eq:orthogonal-GW}
\end{align}
Comparing Eq.~(\ref{Eq:schrodinger+delta}) and Eq.~(\ref{Eq:schrodinger-DFT}), we obtain a linear correction on energy due to the NL potential:
\begin{equation}
\label{Eq:GW-correction}
E_{n}=\xi^{n}+\D_{n}.
\end{equation}
The XC potential can be calculated from the XC free energy, following the discussion in the main text:
\begin{align}
\label{Eq:dFdchinnss}
\frac{\partial F_{\rm XC}}{\partial \chi_{n}}=\D_{n }
\end{align}
\\
\indent
To compute the XC free energy, we consider the Feynman diagram in Fig.~\ref{Fig:Fey}(d) where the phonon Green's function is replaced by the Coulomb potential. We will focus on periodic system and utilize the Bloch wave function, by substituting:
\begin{align}
\phi_{n}(\bfr)\rightarrow \psi_{n\bfk}(\bfr)=\sum_{\bfG}u_{n\bfk}^{\bfG}e^{i(\bfG+\bfk)\bfr}
\end{align}
where $\bfG$ is the unit vector in reciprocal lattice. Thus, we are able to write down the Green's function using the Matsubara representation for a thermal system:
\begin{align}
&G(\bfr,\bfr';\w_n)=-\int^\b_0 d\tau e^{i\w_n\tau}
\<\mathcal{T}\Psi(\bfr,\tau)\Psi^\dagger(\bfr',0)\>\nn\\
&=\sum_{n\bfk}\frac{\psi_{n\bfk}(\bfr)\psi^*_{n\bfk}(\bfr')}{i\w_n-E_{n\bfk}}
=\sum_{n\bfk}G_{n\bfk}(\bfr,\bfr';\w_n)
\end{align}
with $\w_n=(2n+1)\pi/\b$ being the Matsubara frequency. And we can compute the diagram by the Feynman rule:
\begin{align}
&F^d_{\rm col.}=\sum_{\substack{n,m,\bfk,\bfq\\\w_n,\v_n}}
\int \dr \drp  W(\bfq,\v_n-\w_n,\bfr',\bfr)
\nn\\
&~~~~~~~~~~~~~~~~~~~~
\times G_{m\bfk-\bfq}(\bfr,\bfr';\w_n)G_{n\bfk}(\bfr',\bfr;\v_n)
\nn\\
&=\sum \frac{\e^{-1}_{\bfG\bfG'}(\bfq,\w_n-\v_n)
\rho^\bfG_{n m}(\bfk,\bfq)
\rho^{\bfG'*}_{n m}(\bfk,\bfq)}
{(i\w_n-E_{n\bfk})(i\v_n-E_{m\bfk-\bfq})|\bfq+\bfG||\bfq+\bfG'|}
\end{align}
where we define the dipole element:
\begin{equation}
\rho^\bfG_{n m}(\bfk,\bfq)=
\sum_{\bfG_1}u^{\bfG_1+\bfG*}_{n\bfk}
u^{\bfG_1}_{m\bfk-\bfq}
\end{equation}
The dynamical Coulomb screening can be approximated by the plasmon pole approximation (PPA)  \cite{hybertsen1986electron,godby1989metal,larson2013role} as:
\begin{align}
\label{Eq:PPA}
\e^{-1}_{\bfG\bfG'}(\bfq,\w)&\approx\d_{\bfG\bfG'}+\mathbf{R}_{\bfq,\bfG\bfG'}\nn\\
\times&
\[\frac{1}{\w-\Omega_{\bfq,\bfG\bfG'}+i0^+}-\right.
 \left. \frac{1}{\w+\Omega_{\bfq,\bfG\bfG'}-i0^+}\]
\end{align}
where the residue $\mathbf{R}_{\bfq,\bfG\bfG'}$ and the effective plasmon frequency can be obtained by fitting Eq.~(\ref{Eq:PPA}) with the real dielectric constant computed at $\w=0$ and a chosen PPA imaginary frequency $\w=i\w_p$ \cite{godby1989metal}. By summing the Matsubara frequency, we can obtain:

\begin{align}
\label{Eq:Ib}
&F^d=\sum_{\substack{\bfG\bfG' \bfk\bfq nm}}
\frac{
\rho^\bfG_{n m}(\bfk,\bfq)
\rho^{\bfG'*}_{n m}(\bfk,\bfq)}
{|\bfq+\bfG||\bfq+\bfG'|}
\Bigg\{
\d_{\bfG,\bfG'}f_{\bfk,n}f_{\bfk-\bfq,m}
\nn\\
&\left.-\mathbf{R}_{\bfq,\bfG\bfG'}
\[
\frac{f_{\bfk,n}f_{\bfk-\bfq,m}+
f_{\bfk-\bfq,m}\mathcal{N}_{\bfq,\bfG\bfG'}^+
+f_{\bfk,n}\mathcal{N}_{\bfq,\bfG\bfG'}^-}
{\Omega_{\bfq,\bfG\bfG'}-E_{n\bfk}+E_{m\bfk-\bfq}}
\]\right.\nn\\
&+\mathbf{R}_{\bfq,\bfG\bfG'}
\[
\frac{f_{\bfk,n}f_{\bfk-\bfq,m}+
f_{\bfk-\bfq,m}\mathcal{N}_{\bfq,\bfG\bfG'}^-
+f_{\bfk,n}\mathcal{N}_{\bfq,\bfG\bfG'}^+}
{-\Omega_{\bfq,\bfG\bfG'}-E_{n\bfk}+E_{m\bfk-\bfq}}
\]
\Bigg\}.
\end{align}
where we use the shorthand $\mathcal{N}_{\bfq,\bfG\bfG'}^\pm$ for the bosonic occupation $n_{\rm B}(\pm\Omega_{\bfq,\bfG\bfG'})$.
In the limit when the dielectric tensor contains only diagonal term and no frequency dependence, i.e. $\mathbf{R}_{\bfq,\bfG\bfG'}=0$, Eq.~(\ref{Eq:Ib}) reduces to the case without dynamical screening derived in Ref.~\cite{luders2005ab}. Applying Eq.~(\ref{Eq:dFdchinnss}), we can obtain the energy correction from the NL potential:
\begin{align}
&\D_{n\bfk}=
\sum_{\bfG\bfG' \bfq m}
\frac{
\rho^\bfG_{n m}(\bfk,\bfq)
\rho^{\bfG'*}_{n m}(\bfk,\bfq)}
{|\bfq+\bfG||\bfq+\bfG'|}
\times
\nn\\
&\left\{
f_{\bfk-\bfq,m}
\[
\d_{\bfG,\bfG'}+
\frac{2\Omega_{\bfq,\bfG\bfG'}\mathbf{R}_{\bfq,\bfG\bfG'}}{(E_{n\bfk}-E_{m\bfk-\bfq})^2-\Omega_{\bfq,\bfG\bfG'}^2}
\]
\right.\nn\\
&~~\left.+\mathbf{R}_{\bfq,\bfG\bfG'}
\[
\frac{\mathcal{N}_{\bfq,\bfG\bfG'}^-}
{-\Omega_{\bfq,\bfG\bfG'}+E_{n\bfk}-E_{m\bfk-\bfq}}
-
\(\Omega\rightarrow -\Omega\)
\]
\right\}.
\end{align}
This result reproduce the GW correction \cite{hybertsen1986electron} where the first occupation depending term gives the screened exchange part, $\Sigma_{\rm SEX}$, while the second term produce the electron-hole part $\Sigma_{\rm COH}$.

\section{Dressed phonon frequency in effective Hamiltonian Eq.~(\ref{Eq:new-Ha}) and Eq.~(\ref{Eq:new-Hb})\label{Append:self_ph}}
\begin{figure}[t]
    \centering
    \includegraphics[scale=0.19]{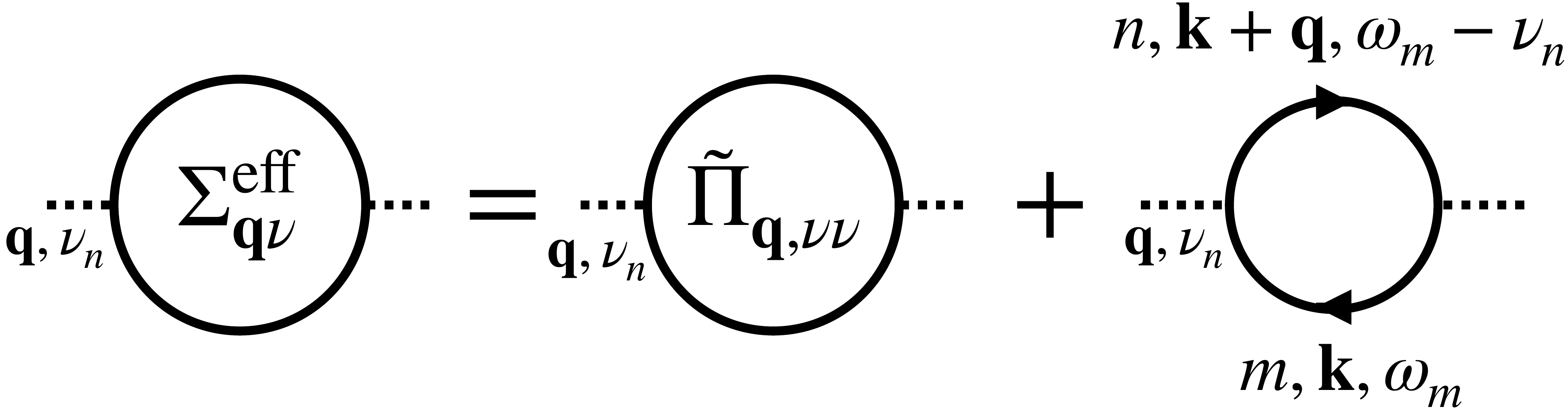}
    \caption{The phonon self-energy diagrams due to the interacting term, Eq.~(\ref{Eq:new-Hb}) where $n,~m$ are the band index for the electron loops and $\w_m$ and $\nu_n$ are the Matsubara frequency of electron and phonon, respectively.}
    \label{Fig:ph_self}
\end{figure}

\vspace{-10pt}
Here we present the detail to derive the renormalized phonon frequency in the diagonal approximation, Eq.~(\ref{Eq:w2-w2}), based on the unperturbed Hamiltonian Eq.~(\ref{Eq:new-Ha}) under the effect of interacting Hamiltonian Eq.~(\ref{Eq:new-Hb}). In terms of Feynman's diagram, the lowest order phonon self-energy can be presented by the one-particle irreducible Feynman's diagrams as in Fig.~\ref{Fig:ph_self} where the first diagram is the effective term from anharmonic ph-ph interaction subtracted by the screened part at zero temperature, and the second term is the contribution from the screening effect at finite temperature.
\\
\indent
Using the Feynman rule, the effective self-energy can be written as:
\begin{align}
    &\Sigma_{\bfq \nu}^{\rm eff}(i\nu_n)
    =\tilde{\Pi}_{\bfq,\nu \nu}\nn\\
    &+2\sum_{nm,\bfk\w_m}|g^{\rm DFPT}_{nm\nu}(\bfk,\bfq)|^2
    \frac{1}{i\w_m-\xi_{\bfk}^m}
    \frac{1}{i(\w_m-\nu_n)-\xi_{\bfk+\bfq}^n}
\end{align}
where the factor of two in the second line appears to take into account the spin degree of freedom. Summing over the Matsubara frequency and taking the analytical continuation to a negligible phonon frequency $i\nu_n \rightarrow \w_{\bfq\nu} \approx 0$, we can obtain:
\begin{equation}
    \Sigma_{\bfq \nu}^{\rm eff}=\tilde{\Pi}_{\bfq,\nu \nu}
    +\frac{X_{\bfq \nu}(T)}{\w^{\rm DFPT}_{\bfq\nu}}. 
\end{equation}
Therefore, the dressed phonon frequency becomes Eq.~(\ref{Eq:w2-w2}):
\begin{align}
    \Omega_{\bfq\nu}^2&=
    {\w^{\rm DFPT}_{ \bfq\v}}^2+2\w^{\rm DFPT}_{ \bfq\v}\Sigma_{\bfq \nu}^{\rm eff}\nn\\
    &=
    {\w^{\rm DFPT}_{ \bfq\v}}^2+2X_{\bfq\v}(T)\Big|^{T}_{T=0}+2\Omega_{\bfq\v}\Sigma_{\bfq\v}(T).
\end{align}
 which is designed to be consistent with the phonon frequency computed by SCP theory.

\section{One-dimnesional two-band model \label{Append:MF}}
\vspace{-10pt}
In this appendix, we work out the MF gap equations Eq.~(\ref{Eq:gap-eq-Db}) and Eq.~(\ref{Eq:gap-eq-Dk}) in the most simplified case in one-dimsional system with two isolated bands. By focusing on a single transition momentum $q={\rm M}=\pi/2$, the system can be described by a reduced Hamiltonian:
\begin{eqnarray}
H&=&\sum_{k} \xi^c_{k} \hat{c}^\dagger_{k}\hat{c}_{k}+\xi^v_{k} \hat{f}^\dagger_{k}\hat{f}_{k}+\Omega \hat{b}^\dagger_{\rm M} \hat{b}_{\rm M}\nn\\
&+&\sum_{k}\(g_k\hat{c}^\dagger_{k+{\rm M}}\hat{f}_{k}+h.c.\)(\hat{b}^\dagger_{-{\rm M}}+\hat{b}_{{\rm M}})
\end{eqnarray}
where we consider only one vibration mode and keep e-ph interaction for phonon with momentum $q= {\rm M}$. $g_k$ is the e-ph matrix element describing the scattering between holes in an occupied state $f$ and electrons in an unoccupied state $c$ by phonon exchanging. In the exciton condensation phase, the electron and hole operator acquires a finite thermal expectation value, $\<\hat{c}^\dagger_{k+{\rm M}}\hat{f}_{k}\>$, accompanied by a finite $\<\hat{b}_{{\rm M}}\>$ for phonon operator such that the electronic ground state becomes unstable and atoms distort from their equilibrium position. Following the procedure introduced in the main text, we first work on Eq.~(\ref{Eq:new-Ha}) by solving the eigenvalue problem:
\begin{align}
&\begin{pmatrix}
\xi^c_{k+{\rm M}}& \D^{cv}_{k}\\
\D^{vc}_{k}& \xi^v_k
\end{pmatrix}
\begin{pmatrix}
t^n_{ck} \\ t^n_{vk}
\end{pmatrix}
=E_{n(k)}
\begin{pmatrix}
t^n_{ck} \\ t^n_{vk}
\end{pmatrix}
\end{align}
which gives the eigenenergy:
\begin{equation}
E_{\pm k}=
\frac{\xi^c_{k+{\rm M}}+\xi^v_{k}}{2}\pm
W_k
\end{equation}
and eigenvectors:
\begin{align}
t^+_{ck}=\frac{e^{i\theta_k}}{\sqrt{2}}
\sqrt{1+\frac{\xi^c_{k+{\rm M}}-\xi^v_k}{2W_k}},~
t^+_{vk}=\frac{1}{\sqrt{2}}\sqrt{1-\frac{\xi^c_{k+{\rm M}}-\xi^v_k}{2W_k}}
\nn\\
t^-_{ck}=\frac{-1}{\sqrt{2}}\sqrt{1-\frac{\xi^c_{k+{\rm M}}-\xi^v_k}{2W_k}},~
t^-_{vk}=\frac{e^{-i\theta_k}}{\sqrt{2}}\sqrt{1+\frac{\xi^c_{k+{\rm M}}-\xi^v_k}{2W_k}}
\end{align}
where 
\begin{equation}
 W_k=\sqrt{\(\frac{\xi^c_{k+{\rm M}}-\xi^v_{k}}{2}\)^2+|\D^{cv}_k|^2}~~;~~
 \theta_k=arg(\D^{cv}_k).
\end{equation}
Based on the solution, we can write down the NL density:
\begin{equation}
\label{Eq:chi_cv_MF}
\chi^{cv}_k=t^{+*}_{ck}t^{+}_{vk}f_{k,+}+t^{-*}_{ck}t^{-}_{vk}f_{k,-}=
\frac{{\D_k^{cv}}^*}{2W_k}(f_{k,+}-f_{k,-}) .
\end{equation}
On the other hand, for the phonon, we take the result, Eq.~(\ref{Eq:b_Db}) from next section:
\begin{equation}
    \label{Eq:chi_b_MF}
    \chi_b=\<b_{\rm M}\>=\frac{-\D_b}{\Omega}.
\end{equation}
As a result, using Eq.~(\ref{Eq:chi_cv_MF}) and Eq.~(\ref{Eq:chi_b_MF}) for Eq.~(\ref{Eq:gap-eq-Db}) and Eq.~(\ref{Eq:gap-eq-Dk}), we can obtain the gap function:
\begin{equation}
    1=\sum_{k}\frac{2|g_k|^2}{\Omega}\frac{f_{k,-}-f_{k,+}}{W_k},
\end{equation}
which reduces to the result in Ref.~\cite{singh2017stable} when we neglect the momentum dependence in e-ph coupling by taking $g_k\equiv g$. 

\section{Phonon Green's function with displacement potential \label{Append:ph_G}}
\vspace{-10pt}
In this section, we derive the phonon Green's function when the phonon operator is applied by an external field $\D_b$. Consider a phonon Hamiltonian:
\begin{equation}
H_0=\Omega b^\dagger b+
\D_b b^\dagger+\D_b^* b
\end{equation}
Due to $\D_b$, the phonon operator has a non-zero expectation value:
\begin{eqnarray}
\label{Eq:b_Db}
\<b\>&=&\frac{1}{Z}{\rm Tr}\[b e^{-\b\Omega(b^\dagger+\D_b^*/\Omega)(b+\D_b/\Omega)+\b |\D_b|^2/\Omega}\]\nn\\
&=&\frac{1}{Z}{\rm Tr}\Big[(b+\D_b/\Omega-\D_b/\Omega)\nn\\
&\times& \exp\[-\b\Omega(b^\dagger+\D_b^*/\Omega)(b+\D_b/\Omega)+\b |\D_b|^2/\Omega\]\Big]\nn\\
&=&\frac{-\D_b}{\Omega}
\end{eqnarray}
where $Z={\rm Tr}e^{-\b H_0}$ is the partition function. Similarly, we can obtain the thermal average of 2-point operators:
\begin{eqnarray}
\<b^\dagger b\>&=&\Bigl<(b^\dagger+\frac{\D_b^*}{\Omega})( b+\frac{\D_b}{\Omega})-\frac{\D_b^*b+\D_b b^\dagger}{\Omega}-\frac{|\D_b|^2}{\Omega^2}\Bigr>\nn\\
&=&n_b(\Omega)+\frac{|\D_b|^2}{\Omega^2}
\end{eqnarray}
and
\begin{eqnarray}
\<b^\dagger b^\dagger\>&=&
\Bigl<(b^\dagger+\frac{\D_b^*}{\Omega})
  (b^\dagger+\frac{\D_b^*}{\Omega})-\frac{2\D^*_b b^\dagger}{\Omega}-\frac{{\D_b^*}^2}{\Omega^2}\Bigr>
  =\frac{{\D_b^*}^2}{\Omega^2}\nn\\
\<b b\>&=&\frac{{\D_b}^2}{\Omega^2}.
\end{eqnarray}
Based on these relations, we can compute the phonon Green's function:
\begin{eqnarray}
D(\tau)\equiv-\<T_\tau A(\tau)A(0)\>
\end{eqnarray}
where 
\begin{align}
A(\t)&=&e^{\t\Omega(b^\dagger+\frac{\D_b^*}{\Omega})(b+\frac{\D_b}{\Omega})}(b+b^\dagger)
e^{-\t\Omega(b^\dagger+\frac{\D_b^*}{\Omega})(b+\frac{\D_b}{\Omega})}\nn\\
&=&(b^\dagger+\frac{\D_b^*}{\Omega})e^{\t\Omega}+(b+\frac{\D_b}{\Omega})e^{-\tau\Omega}-\frac{\D_b^*+\D_b}{\Omega}.
\end{align}
Thus for $\t >0$, we have
\begin{eqnarray}
&&D(\tau)=\nn\\
&&-\<\((b^\dagger+\frac{\D_b^*}{\Omega})e^{\t\Omega}+(b+\frac{\D_b}{\Omega})e^{-\tau\Omega}-\frac{\D_b^*+\D_b}{\Omega}\)(b^\dagger+b)\>=\nn\\
&&-\<(b^\dagger+\frac{\D_b^*}{\Omega})e^{\t\Omega}b+(b+\frac{\D_b}{\Omega})e^{-\tau\Omega}b^\dagger\>+\frac{(\D_b^*+\D_b)\<b^\dagger+b\>}{\Omega}=\nn\\
&&-\(n_B(\Omega)e^{\t\Omega}+(1+n_B(\Omega))e^{-\t\Omega}\)-\frac{(\D_b^*+\D_b)^2}{\Omega^2}
\end{eqnarray}
such that we can obtain the Green's function in terms of Matsubara frequency by the Fourier transformation:
\begin{eqnarray}
\label{Eq:D}
&&\tilde{D}(i\w_n)=\int_0^\beta d\tau e^{i\w_n\t}D(\tau)\nn\\
&&=\(\frac{1}{i\w_n-\Omega}-\frac{1}{i\w_n+\Omega}\)-\d_{\w_n,0}\frac{\beta(\D_b^*+\D_b)^2}{\Omega^2}.
\end{eqnarray}

\section{Derivation of Eq.~(\ref{Eq:Fddchi}) \label{Append:dUdchi}}
\vspace{-10pt}
In this section, we present the steps to obtain Eq.~(\ref{Eq:Fddchi}). We first apply the perturbation theory to second order such that for a state with eigenenergy $\xi^{v_1}$ we can write down the eigenvector:
\begin{align}
   t^{v_1}_{v_2\bfk}&=
   \d_{v_1v_2}\(1-\frac{1}{2}\sum_{c_1}\frac{|\D^{c_1v_1}_{\bfk}|^2}{|\xi^{v_1}_{\bfk}-\xi^{c_1}_{\bfk+{\bf M}}|^2}\)\nn\\
   &~~~~~~~
   +\sum_{c_1}
   \frac{(1-\d_{v_1v_2}){\D_{\bfk}^{c_1v_2}}^*\D_{\bfk}^{c_1v_1}}{(\xi^{v_1}_{\bfk}-\xi_{\bfk}^{v_2})(\xi^{v_1}_{\bfk}-\xi^{c_1}_{\bfk+{\bf M}})} 
\end{align}
and
\begin{align}
   t^{v_1}_{c_1\bfk}=\frac{\D^{v_1c_1}_{\bfk}}{\xi^{c_1}_{\bfk+{\bf M}}-\xi_{\bfk}^{v_1}}   
\end{align}
where we use the band index for the new state as mentioned in the main text. Therefore, the absolute square contains only the diagonal terms:
\begin{equation}
    |t^{v_1}_{v_2\bfk}|^2=\d_{v_1v_2}\(1-\sum_{c_1}\frac{|\D^{c_1v_1}_{\bfk}|^2}{|\xi_{\bfk}^{v_1}-\xi_{\bfk+{\bf M}}^{c_1}|^2}\)~~;~~|t^{v_1}_{c_1\bfk}|^2=0
\end{equation}
which simplifies the Eq.~(\ref{Eq:Fd}) as:
\begin{eqnarray}
&&U^{(d)}\approx
\nn\\
&&
\sum_{\bfk\bfq\nu}\bigg[\sum_{v_1v_2}
I(\xi^{v_2}_{\bfk+\bfq},\xi^{v_1}_{\bfk},\Omega_{\bfq\nu})
|t^{v_1}_{v_1\bfk}|^2|g_{v_2v_1\nu}(\bfk,\bfq)|^2\nn\\
&&+\sum_{c_1c_2}
I(\xi^{c_2}_{\bfk+\bfq+{\bf M}},\xi^{c_1}_{\bfk+{\bf M}},\Omega_{\bfq\nu})
|t^{c_1}_{c_1\bfk}|^2|g_{c_2c_1\nu}(\bfk+{\bf M},\bfq)|^2
\bigg]
\nn\\
\end{eqnarray}
where we keep the $\D^{cv}$ dependence on the states with momentum $(\bfk)$ and reduce the $t^{v_1}_{v_2\bfk+\bfq}$($t^{c_1}_{c_2\bfk+\bfq}$) to the Kronecker delta function. Now we can compute the derivative respective to $|\D^{cv}_{\bfk}|$:
\begin{align}
    &\frac{\partial U^{(d)}}{\partial |\D^{cv}_{\bfk}|}=
    \frac{-2|\D^{cv}_{\bfk}|}{(\xi^c_{\bfk}-\xi^v_{\bfk})^2}
    \nn\\
& ~~~~~~\sum_{\bfq\v}\bigg[\sum_{v_2}
I(\xi^{v_2}_{\bfk+\bfq},\xi^{v_1}_{\bfk},\Omega_{\bfq\nu})
|g_{v_2v\nu}(\bfk,\bfq)|^2\nn\\
&~~~~~~+\sum_{c_2}
I(\xi^{c_2}_{\bfk+\bfq+{\bf M}},\xi^{c_1}_{\bfk+{\bf M}},\Omega_{\bfq\nu})
|g_{c_2c\nu}(\bfk+{\bf M},\bfq)|^2
\bigg].
\end{align}
To change the variable from $\D^{cv}$ to $\chi^{cv}$, we apply the small $\D^{cv}$ expansion to the definition Eq.~(\ref{Eq:NL-chi_b}) such that in the lowest order they follow the relation:
\begin{equation}
    \chi^{cv}_{\bfk}=\frac{-\D^{cv}_{\bfk}}{\xi^c_{\bfk+{\bf M}}-\xi^v_{\bfk}}\(f_{\bfk+{\bf M},c}-f_{\bfk,v}\).
\end{equation}
Therefore, we can use the chain rule to derive the derivative with respect to $\chi^{cv}$ and get Eq.~(\ref{Eq:Fddchi}):
\begin{align}
&\frac{\partial U^{(d)}}{\partial \chi_{\bfk}^{cv*}}
=\frac{\D^{cv}_{\bfk}}{(\xi^{c}_{\bfk+{\bf M}}-\xi^{v}_{\bfk})
\(f_{\bfk+{\bf M},c}-f_{\bfk,v}\)}\nn\\
&~~~~~\times\sum_{\bfq\nu}\bigg[\sum_{v_2}I(\xi^{v_2}_{\bfk+\bfq},\xi^{v}_{\bfk},\w_{\bfq\nu}) |g_{v_2v\nu}(\bfk,\bfq)|^2\nn\\
&~~~~~~~~~~~~~+\sum_{c_2}
I(\xi^{c_2}_{\bfk+{\bf M}+\bfq},\xi^{c}_{\bfk+{\bf M}},\w_{\bfq\nu}) |g_{c_2c\nu}(\bfk+{\bf M},\bfq)|^2\bigg].\nn\\
\end{align}
\section{Numerical detail \label{Append:numerical}}
\vspace{-10pt}
For the DFT and DFPT calculations, we employ a $\Gamma$-centered $36\times 36\times1$ k-grid for BZ sampling and a 85 Ry kinetic energy cutoff for electron density in the self-consistent (scf) calculation with spin-orbit coupling (SOC). By keeping the interlayer vacuum $c=12.5 ~\AA$ to mimic single-layer system, we obtained a relaxed lattice constant $a=3.545~\AA$ in agreement with experimental observation $a_{\rm exp.}=3.538~\AA$ \cite{peng2015molecular}. For the GW correction, based on an alternative scf calculation, we adopt a $20\times 20\times1$ k-grid and 96 real frequency points to sample the dynamical screening. We conduct the self-consistent GW0 calculation with 200 empty bands included and reach convergence in four iterations. For the anharmonic SCP, we construct a $6\times 6 \times 1$ supercell to compute the interatomic force without SOC and use the least absolute shrinkage and selection operator (LASSO) method to extract the anharmonic ph-ph coupling constant. Last, the gap equation is computed with six conduction bands and six valance bands on a $60\times 60\times1$ k-grid with SOC.

\bibliographystyle{apsrev4-2}
\bibliography{TiSe2_ref}
\end{document}